\documentclass[sts]{imsart}

\RequirePackage{amsthm,amsmath,amsfonts,amssymb}
\RequirePackage[authoryear]{natbib}
\RequirePackage{graphicx}
\usepackage{algorithm}
\usepackage{algpseudocode}
\usepackage{dsfont}
 \usepackage{booktabs}
 \usepackage{url} 
  \usepackage{multirow} 
  \usepackage{color}
\usepackage{hyperref}

\newcommand{\E}{\mathbb{E}}

\startlocaldefs
\theoremstyle{plain}

\newtheorem{theorem}{Theorem}[section]
\newtheorem{lemma}[theorem]{Lemma}
\newtheorem{corollary}{Corollary}
\theoremstyle{definition}

\newtheorem{remark}{Remark}
\endlocaldefs

\begin{document}

\begin{frontmatter}

\title{A Calibration Framework for Inference with Partially Observed Data
}
\runtitle{Calibration with partially observed data}

\begin{aug}
\author[A]{\fnms{Mst Moushumi}~\snm{Pervin}\ead[label=e1]{moushumi@iastate.edu}}
\author[B]{\fnms{Hengfang}~\snm{Wang}\ead[label=e2]{hengfang@fjnu.edu.cn}}
\author[C]{\fnms{Jae Kwang}~\snm{Kim}\ead[label=e3]{jkim@iastate.edu}\orcid{0000-0002-0246-6029}
}

\address[A]{Mst Moushumi Pervin is PhD student, Department of Statistics, Iowa State University, Ames, USA\printead[presep={\ }]{e1}.}
\address[B]{Hengfang Wang is Assistant Professor, School of Mathematics and Statistics, Fujian Normal University\printead[presep={\ }]{e2}.}
\address[C]{Jae Kwang Kim is Professor, Department of Statistics, Iowa State University, Ames, USA\printead[presep={\ }]{e3}.}
\end{aug}
\begin{abstract}
Many problems that are often studied separately, including causal inference, semi-supervised learning, and regression with incomplete covariates, share a common inferential structure: the parameter of interest is defined through an estimating equation involving a full-data vector, while part of that vector is unobserved. We study such problems under missing at random through generalized entropy calibration.

Our central claim is that calibration can be understood not only as a device for balancing covariates or stabilizing weights, but also as a general representation of augmentation. The proposed method constructs weights by minimizing a convex entropy subject to a balancing constraint, which incorporates outcome-related information through a calibration function, and a debiasing constraint, which is induced by a working propensity-score model.

This perspective yields a doubly robust estimator that attains the semiparametric efficiency bound when both working models are correct and, under correct propensity-score specification, can be more efficient than classical augmented inverse-probability weighting when the outcome model is misspecified. The framework also accommodates cross-fitting and applies across several partially observed-data problems within a common estimating-equation formulation. Simulations and real-data analyses illustrate the practical consequences of the theory.
\end{abstract}

\begin{keyword}
\kwd{Partially observed data}
\kwd{Calibration}
\kwd{Doubly robust estimation}
\kwd{Generalized entropy calibration}
\kwd{Semi-supervised learning}
\kwd{Causal inference}
\end{keyword}

\end{frontmatter}
\newpage

\section{Introduction}\label{sec:intro}

Many problems that are usually studied in separate literatures can be viewed as instances of a common inferential problem. In causal inference, one of the potential outcomes is unobserved for each unit. In semi-supervised learning, responses are observed only for a labeled subset. In regression with incomplete covariates, part of the predictor vector is missing. Although these settings differ in scientific interpretation, they share the same statistical structure: the parameter of interest is defined through a full-data estimating equation, but part of the full-data vector is unobserved. From this viewpoint, they are all problems of inference with partially observed estimating equations under missing at random (MAR); see, for example, \citet{rubin1976}, \citet{robins1994estimation}, \citet{tsiatis2006}, \citet{little2019statistical}, and \citet{kimshao2021}.

The methodological responses to this problem have developed in different languages. In missing-data inference, the dominant themes have been outcome regression, inverse-probability weighting, and doubly robust augmentation \citep{robins1994estimation, rotnitzky1995, scharfstein1999, bang2005, cao2009, tsiatis2006}. In survey sampling and nonresponse adjustment, calibration weighting provides a different formulation \citep{deville1992, deville2000, wu2001}. 
In causal inference, balancing weights have become a central tool for covariate adjustment \citep{hainmueller2012, imai2014, li2018balancing, zhao2019, tan2020regularized, fan2022optimal, hirshberg2021augmented, benmichael2021augmented}. Related work has also emphasized the close connection between balancing and modeling perspectives on weighting \citep{chattopadhyay2020balancing}.
Semi-supervised learning has developed its own efficiency-based perspective on the use of unlabeled data \citep{chakrabortty2018efficient, gronsbell2018semi, zhang2019semi, azriel2022semi}. What remains less clear is how these strands fit together within a single estimating-equation framework.

A standard route through this problem starts from two familiar ideas. Outcome-regression (OR) methods replace the missing part of the data by a conditional mean or prediction, whereas propensity-score (PS) weighting corrects selective observation by reweighting the complete cases using inverse selection probabilities \citep{horvitz1952}. Doubly robust (DR) procedures combine these two strategies and remain consistent when either the OR model or the PS model is correctly specified \citep{robins1994estimation, rotnitzky1995, scharfstein1999, bang2005, cao2009}. This logic is usually written in additive form: one starts from an inverse-probability-weighted estimating equation and then adds an augmentation term. That representation is natural, but it is not the only one.

Our central claim is that augmentation can be represented implicitly through calibration. Rather than adding an explicit augmentation term, we construct weights by minimizing a generalized entropy subject to two constraints: a balancing constraint that incorporates outcome-related information through a calibration function, and a debiasing constraint induced by a working PS model. In this formulation, calibration is not merely a tool for moment balancing or weight stabilization. It becomes a general representation of augmentation itself. This places weighting, balancing, and doubly robust correction within a single framework for inference with partially observed estimating equations.

The present work builds on the generalized entropy calibration framework of \citet{kwon2025debiased}, which was developed for estimating finite-population totals under design-based inference with known auxiliary variables. Our setting is different in both target and interpretation. Here the parameter is defined through a general estimating equation under MAR, the calibration function may depend on the target parameter itself, and estimation proceeds through profile optimization. These differences allow calibration to recover outcome-related information from the full sample while preserving the debiasing role of inverse-probability weighting. Recent work by \citet{brunssmith2025} also highlights links between balancing weights and regression adjustment in linear settings; the present paper develops a broader calibration-based perspective for general estimating equations with partially observed data.

This perspective clarifies the geometry behind double robustness and efficiency. When the calibration function captures the relevant conditional mean structure, the balancing constraint plays the role usually assigned to augmentation. When the PS model is correctly specified, the debiasing constraint recovers the inverse-probability-weighting pathway. Double robustness then emerges from the interaction of these two constraint systems. The same perspective also clarifies why efficiency can improve relative to classical augmented inverse-probability weighting (AIPW). Standard AIPW augments through a working OR term, whereas the generalized entropy calibration estimator projects onto a larger space generated jointly by outcome-related and debiasing covariates. As a result, when the PS model is correct but the working OR model is misspecified, the proposed estimator can have strictly smaller asymptotic variance than classical AIPW, while under joint model correctness it attains the same semiparametric efficiency bound identified by \citet{robins1994estimation}.

The same framework applies naturally across problems that are often treated as distinct. In causal inference, the partially observed component is a counterfactual outcome. In semi-supervised learning, it is an unlabeled response. In regression with incomplete covariates, it is a missing part of the predictor vector. The scientific meanings differ, but the inferential template is the same: one observes an always-observed component, a selectively observed component, and an estimating function defined on the full data. The examples in this paper are included not simply to demonstrate the reach of a new estimator, but to show that the same calibration logic recurs across several domains of modern statistics.

The framework also connects naturally to modern prediction-based inference. Because the balancing constraint requires calibration-function values for all units, including those with unobserved components, it is natural to estimate these quantities using cross-fitted predictions. This links the present approach to double machine learning \citep{chernozhukov2018double} and the recent prediction-powered perspective of \citet{angelopoulos2023prediction}. Here, however, machine learning plays a supporting role: it enriches the calibration space without replacing the underlying estimating-equation structure.

This paper makes three contributions. First, it formulates causal inference, semi-supervised learning, and regression with incomplete covariates within a common framework of inference with partially observed estimating equations under MAR. Second, it shows that calibration can be understood not only as a weighting device but as a general representation of augmentation, thereby linking calibration weighting to the logic of doubly robust inference. Third, it provides a geometric explanation of why this representation yields double robustness, recovers local efficiency under joint model correctness, and can improve upon classical AIPW when the PS model is correct but the working OR model is misspecified.

The remainder of the paper is organized as follows. Section~\ref{sec:prelim} introduces the general setup and reviews OR, IPW, and AIPW. Section~\ref{sec:proposed} presents generalized entropy calibration as a representation of augmentation. Section~\ref{sec:theory} develops the main asymptotic implications of this viewpoint. Section~\ref{sec:computation} discusses computation and cross-fitting. Section~\ref{sec:applications} shows how the framework specializes to causal inference, semi-supervised learning, and regression with missing covariates. Section~\ref{sec:simulation} reports simulation results, and Section~\ref{sec:realdata} presents real-data applications. Some concluding remarks are given in Section~\ref{sec:conclusion}. All technical proofs are deferred to the supplementary material.

\section{Basic Setup}\label{sec:prelim}

We consider a general partially observed-data setting in which the full data vector for unit \(i\) is \(Z_i=(O_i^\top,M_i^\top)^\top\), where \(O_i\) denotes the subvector of \(Z_i\) that is always observed and \(M_i\) denotes the subvector subject to missingness. Let \(\delta_i\in\{0,1\}\) indicate whether \(M_i\) is observed, so that the observed data are \((O_i,\delta_i,\delta_i M_i)\). The parameter of interest \(\theta\in\mathbb{R}^p\) is defined through the full-data estimating equation
\[
\mathbb{E}\{U(\theta;Z_i)\}=0,
\]
where \(U(\theta;Z_i)\) is a known estimating function. This setup encompasses a wide range of problems, including causal inference, semi-supervised learning, and regression with incomplete covariates. The statistical task is to recover the full-data estimating equation, and hence \(\theta\), from partially observed data under assumptions on the missingness mechanism.

To identify \(\theta\) from the observed data, we assume that the missingness mechanism is missing at random (MAR), that is,
\[
\delta_i \perp M_i \mid O_i.
\]
Thus, conditional on the always-observed subvector \(O_i\), the indicator of whether \(M_i\) is observed does not depend further on the missing subvector itself. Let
\[
\pi(O_i)=P(\delta_i=1\mid O_i)
\]
denote the propensity score, and assume positivity, \(0<\pi(O_i)\le 1\) almost surely. Under these conditions, the full-data estimating equation can be recovered from the observed data through inverse-probability weighting, since
\[
\mathbb{E}\!\left\{\frac{\delta_i}{\pi(O_i)}\,U(\theta;Z_i)\right\}
=
\mathbb{E}\{U(\theta;Z_i)\}.
\]
This identity underlies the IPW estimator and also serves as a starting point for augmented and calibration-based constructions.

Under MAR, a natural approach is inverse-probability weighting (IPW), which reconstructs the full-data estimating equation by reweighting the complete observations. Specifically,
\[
\mathbb{E}\!\left\{\frac{\delta_i}{\pi(O_i)}\,U(\theta;Z_i)\right\}
=
\mathbb{E}\{U(\theta;Z_i)\}
\]
suggests estimating \(\theta\) by solving
\[
\frac{1}{N}\sum_{i=1}^N \frac{\delta_i}{\pi(O_i;\hat\phi)}\,U(\theta;Z_i)=0,
\]
where \(\pi(O_i;\phi)\) is a working propensity-score model and \(\hat\phi\) is an estimator of \(\phi\). If the propensity-score model is correctly specified, the resulting IPW estimator is consistent for \(\theta_0\). However, because IPW relies only on the response mechanism, it can be inefficient and numerically unstable when some estimated propensity scores are small. This limitation motivates augmented constructions that incorporate outcome-related information in addition to inverse-probability weighting.

A standard refinement of IPW is augmented inverse-probability weighting (AIPW), which combines weighting with an explicit outcome-regression adjustment. Let \(b(\theta;O_i)\) denote a working model for
\[
b^*(\theta;O_i)=\mathbb{E}\{U(\theta;Z_i)\mid O_i\},
\]
the conditional mean of the full-data estimating function given the always-observed subvector \(O_i\). The AIPW estimator is obtained by solving
\[
\frac{1}{N}\sum_{i=1}^N \left[
b(\theta;O_i)
+
\frac{\delta_i}{\pi(O_i;\hat\phi)}
\bigl\{U(\theta;Z_i)-b(\theta;O_i)\bigr\}
\right]=0.
\]
This construction is doubly robust: it remains consistent for \(\theta_0\) if either the propensity-score model or the outcome-regression model is correctly specified. In addition, when both working models are correct and \(b(\theta;O_i)=b^*(\theta;O_i)\), the estimator attains the semiparametric efficiency bound. Thus, AIPW improves upon IPW by incorporating outcome-related information through an explicit augmentation term 
\citep{robins1994estimation}.

However, the AIPW estimator relies on an explicit augmentation term $b(\theta; O_i)$ and can be sensitive to extreme inverse-probability weights, motivating the calibration-based approach developed in Section~\ref{sec:proposed}.

\section{Proposed Estimators}\label{sec:proposed}

The augmented inverse-probability weighted estimator improves upon IPW by adding an explicit augmentation term $b(O_i)$ to the estimating equation. Our starting point is that this augmentation step can instead be represented implicitly through calibration. Rather than inserting $b(O_i)$ directly into the estimating equation, we construct weights so that the weighted complete cases reproduce both outcome-related information and response-mechanism information. This leads to a calibration-based formulation of doubly robust estimation in which weighting, augmentation, and balance are integrated within a single framework.

More specifically, the proposed weights are designed to satisfy two goals. The first is to balance a calibration function
\[
b(O)=\bigl(1,\tilde b(O)^\top\bigr)^\top
\]
that approximates the conditional mean
\[
b^*(\theta;O)=E\{U(\theta;Z)\mid O\},
\]
thereby capturing the same information that appears in the classical augmentation term. We assume throughout that the first component of \(b(O)\) is an intercept term. Consequently, the balancing constraint enforces not only moment balance for the nonconstant components of \(b(O)\), but also total-weight calibration. The second goal is to incorporate a working propensity-score model through an additional debiasing constraint, so that the resulting weights recover the inverse-probability weighting structure when the PS model is correctly specified. In this sense, the proposed construction provides a calibration representation of the usual OR/IPW synthesis.

We therefore define the estimator $\hat\theta_\omega$ as the solution to the weighted estimating equation
\begin{equation}\label{eq:weighted_ee}
  \sum_{i=1}^{N} \delta_i\, \hat\omega_i(\theta)\, U(\theta; z_i) = 0,
\end{equation}
where $\hat\omega_i(\theta)=\hat\omega(\theta;O_i)$ denotes a calibration weight assigned to unit $i$. The dependence on $\theta$ is essential: because the target estimating function itself depends on $\theta$, the auxiliary information used to calibrate the weights may also depend on $\theta$.

To determine $\hat\omega_i(\theta)$, we employ the generalized entropy calibration method of \citet{kwon2025debiased}. Let $G\colon \nu \to \mathbb{R}$ be a strictly convex, differentiable function with derivative $g=G'$. The calibration weights for the observed units ($\delta_i=1$) are obtained by solving
\begin{equation}\label{eq:primal}
  \min_{\omega_1,\ldots,\omega_N}\sum_{i=1}^{N}\delta_i\,G(\omega_i),
\end{equation}
subject to two constraints, each corresponding to one side of the doubly robust construction.

The first is the \emph{balancing constraint}
\begin{equation}\label{eq:balance}
  \sum_{i=1}^{N}\delta_i\,\omega_i\,  b(O_i)=\sum_{i=1}^{N} b(O_i) ,
\end{equation}
which requires the weighted complete cases to reproduce the empirical moments of the calibration function \(b(O)\). Because the first component of \(b(O)\) is an intercept, \eqref{eq:balance} also implies the total-weight calibration identity
\[
\sum_{i=1}^{N}\delta_i\,\omega_i = N.
\]
When \(b(O)\) approximates \(b^*(\theta;O)=\mathbb{E}\{U(\theta;Z)\mid O\}\), this constraint imports outcome-related information from the full sample into the respondent sample and plays the same role as the augmentation term in AIPW.

The second is the \emph{debiasing constraint}
\begin{equation}\label{eq:debias}
  \sum_{i=1}^{N}\delta_i\,\omega_i\,g(\hat\pi_i^{-1})
  =\sum_{i=1}^{N} g(\hat\pi_i^{-1}),
\end{equation}
where $\hat\pi_i=\pi(O_i;\hat\phi)$ is the fitted PS model. This constraint incorporates information from the response mechanism and anchors the calibration weights to the IPW solution when the PS model is correctly specified. The appearance of $g(\hat\pi_i^{-1})$ is dictated by the entropy geometry. Since $g=G'$, the optimal weights from \eqref{eq:primal} take the form
\[
\omega_i^*=g^{-1}(\lambda^\top s_i),
\]
for an appropriate dual variable $\lambda$; see Section~\ref{sec:dual}. Consequently, if calibration is imposed only through the debiasing constraint and the PS model is correct, then
\[
\omega_i^*=g^{-1}\{g(\pi_i^{-1})\}=\pi_i^{-1},
\]
which exactly recovers the usual IPW weights. Thus, the debiasing covariate $g(\hat\pi_i^{-1})$ is not an ad hoc choice; it is the form required for the calibration weights to inherit the IPW structure under a correct PS model.

The two constraints therefore play distinct but complementary roles. The balancing constraint injects outcome-related information through the calibration function $b(O)$, whereas the debiasing constraint incorporates the fitted propensity-score model and protects consistency under correct PS specification. Together they provide a calibration-based analogue of the classical doubly robust construction. In particular, if $b(O)$ equals $b^*(\theta;O)$, or is consistently estimated from a correct OR model, then balancing removes the leading bias term even when the PS model is misspecified. Conversely, if the PS model is correct, the debiasing constraint ensures that the calibration estimator inherits the consistency of IPW even when the OR model is misspecified. These properties are formalized in Section~\ref{sec:theory}.

Different choices of the entropy function $G$ produce different weighting schemes; several examples are summarized in Table~\ref{tab:entropy}.

\begin{table*}
%\centering
\caption{Examples of generalized entropies $G(\omega)$, the corresponding calibration covariates $\hat g_i = g(\hat\pi_i^{-1})$ and $\hat g_i^{-1}$.}
\label{tab:entropy}
\begin{tabular}{lcccc}
\hline
Entropy & $G(\omega)$ & $\hat g_i$ & $\hat g_i^{-1}$ & Domain \\
\hline
Squared loss & $\omega ^2/2$ & $\hat\pi_i^{-1}$ & $\hat\pi_i$ & $(-\infty, \infty)$ \\
Empirical likelihood & $-\log\omega $ & $-\hat\pi_i$ & $-1/\hat\pi_i$ & $(0, \infty)$ \\
Exponential tilting & $\omega\,\log\omega - \omega$ & $\log(\hat\pi^{-1}_i)$ & $1/\log(\hat\pi^{-1}_i)$ & $(0, \infty)$ \\
Hellinger distance & $-\sqrt{\omega}$ & $-\sqrt{\hat\pi_i/2}$ & $-\sqrt{2/\hat\pi_i}$ & $(0, \infty)$ \\
\hline
\end{tabular}
\end{table*}

\subsection{Dual formulation}\label{sec:dual}

The primal program in \eqref{eq:primal}--\eqref{eq:debias} expresses the statistical idea of the method, but the resulting weights are most conveniently characterized through its dual. The dual formulation serves two purposes. First, it yields a closed-form representation of the optimal weights in terms of a low-dimensional multiplier $\lambda$, making clear how the balancing and debiasing constraints jointly determine the final calibration rule. Second, it reduces the constrained optimization over $N$ weights to an unconstrained convex optimization over $\lambda \in \mathbb{R}^{q+1}$, which is the key computational advantage of generalized entropy calibration.

To derive the dual representation, introduce Lagrange multipliers $\lambda_1 \in \mathbb{R}^q$ and $\lambda_2 \in \mathbb{R}$ for constraints \eqref{eq:balance}--\eqref{eq:debias}, where \(q\) denotes the dimension of \(b(O)\), including its intercept term. Write \(b_i=b(O_i)\), \(\hat g_i=g(\hat\pi_i^{-1})\), and \(s_i=(b_i^\top,\hat g_i)^\top\). The Lagrangian is
\begin{align}
Q(\omega, \lambda)
&= - \sum_{i=1}^N \delta_i\, G(\omega_i)
+ \lambda_1^{\top}
\left( \sum_{i=1}^N \delta_i \,\omega_i \, b_i - \sum_{i=1}^N  b_i \right)\nonumber \\
&\qquad + \lambda_2
\left( \sum_{i=1}^N \delta_i \,\omega_i \,\hat g_i - \sum_{i=1}^N \hat g_i \right),
\label{eq:lag}
\end{align}
where $\lambda = (\lambda_1^\top, \lambda_2)^\top$.

Maximizing $Q(\omega,\lambda)$ with respect to $\omega_i$ yields the closed-form optimal weights
\begin{equation}\label{eq:opt_weights}
  \omega_i^*(\lambda) = g^{-1}(\lambda_1^\top b_i + \lambda_2 \hat g_i)
  = g^{-1}(\lambda^\top s_i),
\end{equation}
where $g^{-1}$ is strictly increasing because $G$ is strictly convex. The function $g ( \cdot)$ is called the calibration link function and is closely related to the canonical link function in the generalized linear models. The calibration link operates on the weight parameter~$\omega_i$ rather than a conditional mean $\mu_i$, but the algebraic structure is identical. 

Substituting \eqref{eq:opt_weights} back to (\ref{eq:lag}) 
gives the dual objective
\begin{equation}\label{eq:dual}
  \rho_G(\lambda) = \frac{1}{N}\sum_{i=1}^{N} \delta_i\, F(\lambda^\top s_i) - \frac{1}{N}\sum_{i=1}^{N} \lambda^\top s_i,
\end{equation}
where $F(\nu) = -G\{g^{-1}(\nu)\} + g^{-1}(\nu)\,\nu$ is the convex conjugate of~$G$.
The calibration step is therefore reduced to the convex optimization problem
\[
\hat\lambda = \arg\min_{\lambda}\rho_G(\lambda),
\]
after which the final weights are obtained as $\hat\omega_i=\omega_i^*(\hat\lambda)$.

Differentiating \eqref{eq:dual} confirms that $\hat\lambda$ satisfies the calibration equations \eqref{eq:balance}--\eqref{eq:debias}:
\[
  \frac{\partial}{\partial \lambda}\rho_G(\lambda)
  = \frac{1}{N}\left\{\sum_{i=1}^{N} \delta_i\, \omega_i^*(\lambda)\, s_i - \sum_{i=1}^{N} s_i\right\}.
\]
 For this reason,  we refer to $\rho_G( \lambda)$ as the calibration generating function induced by the entropy $G$.  The dual representation reduces the problem to optimizing over $\lambda \in \mathbb{R}^{q+1}$, regardless of the sample size~$N$.

\section{Large-sample properties}\label{sec:theory}

This section establishes the large-sample behavior of the proposed estimator $\hat{\theta}_\omega$ and clarifies how the calibration construction reproduces the classical doubly robust logic. The analysis proceeds in three steps. We first derive a first-order linearization showing that the estimator admits a calibration-based augmentation representation without requiring either working model to be correct. We then study the two one-model regimes separately: correctness of the propensity-score model and correctness of the outcome-regression model. These results together imply double robustness. Finally, we compare the proposed estimator with the classical AIPW estimator and show that, under a correct propensity-score model, generalized entropy calibration is never less efficient and is strictly more efficient whenever the debiasing covariate contributes information beyond the augmentation space used by AIPW.

Throughout, $\hat{\theta}_\omega$ solves
\[
\hat{U}^\omega(\theta) \;\equiv\; \sum_{i=1}^{N} \delta_i\, \omega_i^*(\hat{\lambda},\hat{\varphi})\, U(\theta;\, z_i) \;=\; 0,
\]
where $\omega_i^*(\lambda,\varphi) = g^{-1}\!\bigl(\lambda_1^\top b_i + \lambda_2\, g_i(\varphi)\bigr)$, $b_i = b(O_i)$, and $g_i(\varphi) = g\{\pi^{-1}(O_i;\varphi)\}$. The parameters $\hat{\varphi}$ and $\hat{\lambda}$ are obtained jointly from
\begin{align}
\nabla \ell(\varphi) &\;\equiv\; \frac{1}{N}\sum_{i=1}^{N} \biggl(\frac{\delta_i}{\pi_i} - 1\biggr) h_i(\varphi) \;=\; 0, \label{eq:ps-score}\\[4pt]
\nabla \hat{\rho}_G(\lambda) &\;\equiv\; \frac{1}{N}\sum_{i=1}^{N} \Biggl\{\delta_i\, \omega_i^*(\lambda,\hat{\varphi}) \begin{pmatrix} b_i \\ g_i(\hat{\varphi}) \end{pmatrix} - \begin{pmatrix} b_i \\ g_i(\hat{\varphi}) \end{pmatrix}\Biggr\} \;=\; 0, \label{eq:cal-score}
\end{align}
where $\pi_i = \pi(O_i;\varphi)$ and $h_i(\varphi) = \{1 - \pi_i(\varphi)\}^{-1}\, \partial\pi_i(\varphi)/\partial\varphi$.

The regularity conditions are collected in the Supplementary Material (SM), Section~A. Assumptions~1--2 ensure the consistency of $\hat{\varphi}$ and $\hat{\lambda}$, established in Lemma~S.1 of the SM. Assumptions~3--4 impose smoothness and nondegeneracy of the limiting dual and propensity-score objectives. In particular, nonsingularity of the Hessian $E\{\nabla^2 \rho_G(\lambda^*)\}$ requires that the calibration covariates $s_i=(b_i^\top,g_i)^\top$ are not collinear among the respondents, which is the natural identifiability condition for entropy calibration. Assumptions~5--7 are standard conditions for joint $M$-estimation of the coupled system defining $\hat{\theta}_\omega$, $\hat{\lambda}$, and $\hat{\varphi}$.

\subsection{Linearization}

The key technical step is a first-order expansion of $\hat U_\omega(\theta)$ showing that the proposed estimator behaves like an augmentation estimator in which the augmentation term is generated by calibration. This representation is the basis for all subsequent results and already reveals the role of the calibration covariates $S(O;\phi^*)$ in separating full-sample information from respondent-only residual variation.

\begin{theorem}[Linearization]\label{thm:lin}
Under Assumptions~1--4,
\[
\hat{U}^\omega(\theta) \;=\; \tilde{U}^\omega(\lambda^*,\hat{\varphi}) \;+\; o_p(N^{-1/2}),
\]
where
\begin{align*}
    \tilde{U}^\omega(\lambda^*,\hat{\varphi}) \;&=\; \frac{1}{N}\sum_{i=1}^{N} \Bigl[\gamma^*\, s_i(\hat{\varphi}) \;+\; \delta_i\, \omega_i^*(\lambda^*,\hat{\varphi})\bigl\{U(\theta;\, z_i) \\&\qquad- \gamma^*\, s_i(\hat{\varphi})\bigr\}\Bigr],
\end{align*}

$s_i(\hat{\varphi}) = \bigl(b_i^\top,\, g(\pi^{-1}(O_i;\hat{\varphi}))\bigr)^\top$, and $\gamma^* \in \mathbb{R}^{q \times (q+1)}$ is the probability limit of $\hat{\gamma}$ satisfying
\[
\sum_{i=1}^{N} \delta_i\, f'\!\bigl(\lambda^{*\top} s_i(\hat{\varphi})\bigr)\, \bigl\{U(\theta;\, z_i) - \gamma\, s_i(\hat{\varphi})\bigr\}\, s_i^\top(\hat{\varphi}) \;=\; 0,
\]
with $f' = (g^{-1})'$. This expansion holds without assuming correctness of either the PS or the OR model.
\end{theorem}

The expansion in Theorem~4.1 has a useful interpretation. The term $\gamma^* s_i(\hat\phi)$ is the population weighted least-squares projection of the full-data estimating function onto the calibration covariates, so it plays the role of an implicit augmentation term. The remaining respondent-only component is the residual after that projection. Thus, the generalized entropy calibration estimator can be viewed asymptotically as an augmentation estimator whose augmentation space is determined by the calibration constraints. This perspective is central for understanding both the doubly robust property and the variance comparison with AIPW developed below.

\subsection{Consistency under the correct PS model}

We begin with the regime in which the propensity-score model is correctly specified. In this case, the debiasing constraint forces the calibration weights to recover the IPW solution, so the proposed estimator inherits the consistency of inverse-probability weighting while potentially benefiting from additional variance reduction through the calibration structure.

\begin{lemma}\label{lem:ps-correct}
If the PS model is correctly specified, i.e., $P(\delta=1\mid O) = \pi(O;\varphi_0)$, then $\varphi^* = \varphi_0$ and $\lambda_1^* \to 0$, $\lambda_2^* \to 1$, so that $\omega_i^*(\lambda^*,\varphi_0) = 1/\pi(O_i;\varphi_0)$.
\end{lemma}

\begin{corollary}[Asymptotic normality under correct PS model]\label{cor:ps}
Under Assumptions~1--7 and a correctly specified PS model,
\[
\sqrt{N}\,(\hat{\theta}_\omega - \theta_0) \;\xrightarrow{d}\; \mathcal{N}\!\bigl(0,\; \tau_1^{-1}\, V_1\, (\tau_1^{-1})^\top\bigr),
\]
where $\tau_1 = \mathbb{E}\bigl\{  \partial U(\theta_0;Z)/ \partial \theta^\top \bigr\}$ and
\begin{align*}
    V_1 \;&=\; \mathrm{Var}\bigl\{U(\theta_0;Z)\bigr\} \;+\; \mathbb{E}\!\Bigl[\Bigl(\frac{1}{\pi(O;\varphi_0)} - 1\Bigr) \\
    &\qquad \Bigl\{ U(\theta_0;Z) - \gamma^* S(O;\varphi_0) - \kappa^* h(O;\varphi_0)\Bigr\}^{\otimes 2} \Bigr],
\end{align*}

with $S(O;\varphi_0) = \bigl(b^\top(O),\, g(\pi^{-1}(O;\varphi_0))\bigr)^\top$ and $B^\otimes = B B^\top$. Here $\kappa^* \in \mathbb{R}^{q \times r}$ is the probability limit of $\hat{\kappa}$ defined by
\begin{align*}\label{eq:kappa}
&\frac{1}{N}\sum_{i=1}^{N} \frac{\partial}{\partial\varphi} \biggl\{\gamma^* s_i(\varphi) + \delta_i\, \omega_i^*(\lambda^*,\varphi)\bigl[U(\theta;\, z_i) - \gamma^* s_i(\varphi)\bigr]\nonumber\\
&\qquad+ \Bigl(1 - \frac{\delta_i}{\pi(O_i;\varphi)}\Bigr)\kappa\, h_i(\varphi)\biggr\} = 0.
\end{align*}
The term $\kappa^* h(O;\varphi_0)$ captures the effect of estimating $\varphi$ on the asymptotic variance; it vanishes when the OR model is also correctly specified.
\end{corollary}

\subsection{Consistency under the correct OR model}

We next consider the complementary regime in which the outcome-regression model is correctly specified. Here the balancing constraint aligns the calibration function with the conditional mean of the estimating function, so the leading bias term is removed even if the propensity-score model is misspecified.

\begin{lemma}\label{lem:or-correct}
If the OR model is correctly specified, i.e., $\mathbb{E}\{U(\theta_0;Z)\mid O\} \in \mathrm{span}\{b(O)\}$, and $b^*(O) = \mathbb{E}\{U(\theta;Z)\mid O\}$ is used in \eqref{eq:balance}, then $\kappa^* = 0$.
\end{lemma}

\begin{corollary}[Asymptotic normality under correct OR model]\label{cor:or}
Under Assumptions~1--7, if $b^*(O) = \mathbb{E}\{U(\theta;Z)\mid O\}$ satisfies $\mathbb{E}\{U(\theta_0;Z)\mid O\} \in \mathrm{span}\{b(O)\}$, then
\[
\sqrt{N}\,(\hat{\theta}_\omega - \theta_0) \;\xrightarrow{d}\; \mathcal{N}\!\bigl(0,\; \tau_1^{-1}\, \bar{V}_1\, (\tau_1^{-1})^\top\bigr),
\]
where
\begin{align*}
    \bar{V}_1 \;&=\; \mathrm{Var}\bigl\{\mathbb{E}[U(\theta_0;Z)\mid O]\bigr\} \\
    &\qquad\;+\; \mathbb{E}\!\Big[\delta\,\bigl\{\omega^*(O;\lambda^*,\varphi^*)\bigr\}^2\, \mathrm{Var}\{U(\theta_0;Z)\mid O\}\Big].
\end{align*}

Hence $\hat{\theta}_\omega$ remains $\sqrt{N}$-consistent and asymptotically normal even when the PS model is misspecified.
\end{corollary}

\subsection{Double robustness, variance dominance, and local efficiency}
\label{sec:discussion}

Corollaries~\ref{cor:ps} and \ref{cor:or} together establish that $\hat{\theta}_\omega$ is
doubly robust: it is consistent for $\theta_0$ whenever either the PS model or the OR model
is correctly specified. Double robustness alone, however, does not distinguish the proposed
estimator from classical AIPW, which shares this property. The more distinctive question is
comparative: once the PS model is correct, does the calibration formulation offer any
efficiency advantage over AIPW? The answer is yes, and the reason is geometric.

To see why, recall from Theorem~\ref{thm:lin} that $\hat{\theta}_\omega$ behaves
asymptotically like an augmentation estimator whose augmentation term is the projection of
$U(\theta_0; Z)$ onto the calibration covariate space
\[
  \mathcal{S}(O;\varphi_0)
  = \mathrm{span}\bigl\{b(O),\;
    g\bigl(\pi^{-1}(O;\varphi_0)\bigr),\;
    h(O;\varphi_0)\bigr\}.
\]
The classical AIPW estimator, by contrast, projects onto the strictly smaller space
$\mathrm{span}\{b(O)\}$ alone. The variance of an augmentation estimator is determined by
the residual after projection: a larger projection space yields a smaller residual, and
hence a smaller variance. This is the content of Corollary~\ref{cor:var-dom} below.

To state the result precisely, define a weighted seminorm on functions of $(O, Z)$ by
\[
  \|f\|_w^2
  = \mathbb{E}\!\left[
      \left(\frac{1}{\pi(O;\varphi_0)} - 1\right) f^{\otimes 2}
    \right],
\]
where the weight $\pi^{-1}(O;\varphi_0) - 1$ is the excess weight assigned to complete
cases under IPW. The variance components of both estimators can be expressed in terms of
this seminorm: $V_3 - V_1$ equals the squared $\|\cdot\|_w$-distance between the
AIPW projection and the GEC projection of $U(\theta_0; Z)$ onto their respective
augmentation spaces. Since $\mathrm{span}\{b(O)\} \subseteq \mathcal{S}(O;\varphi_0)$,
the Pythagorean identity in $\|\cdot\|_w$ gives
\begin{equation}\label{eq:pythag}
  \bigl\|U - \Pi_{\{b\}} U\bigr\|_w^2
  = \bigl\|U - \Pi_{\mathcal{S}} U\bigr\|_w^2
    + \bigl\|\Pi_{\mathcal{S}} U - \Pi_{\{b\}} U\bigr\|_w^2,
\end{equation}
where $\Pi_A$ denotes $\|\cdot\|_w$-projection onto the space $A$, and $U$ abbreviates
$U(\theta_0; Z)$. The second term on the right of~\eqref{eq:pythag} is non-negative,
establishing $V_1 \leq V_3$.

\begin{corollary}[Variance dominance over AIPW]\label{cor:var-dom}
Under the conditions of Corollary~\ref{cor:ps}, let $V_1$ denote the variance component of the proposed estimator $\hat{\theta}_\omega$ and let
\begin{align*}
  V_3 \;&=\; \mathrm{Var}\bigl\{U(\theta_0;Z)\bigr\} \\
  &\qquad \;+\; \mathbb{E}\!\left[\left(\frac{1}{\pi(O;\varphi_0)} - 1\right) \left\{ U(\theta_0;Z) - b(O)\right\}^{\otimes 2} \right]  
\end{align*}
be the corresponding variance component of the classical AIPW estimator. Then
\begin{equation}\label{eq:var-dom}
V_1 \;\leq\; V_3,
\end{equation}
with equality if and only if $g\bigl(\pi^{-1}(O;\varphi_0)\bigr)$ and $h(O;\varphi_0)$ lie in $\mathrm{span}\{b(O)\}$ almost surely.
\end{corollary}

Corollary~3 identifies the main advantage of the calibration approach over classical AIPW. Standard AIPW augments the estimating equation using only the working outcome-regression component $b(O)$. By contrast, the generalized entropy calibration estimator projects onto the larger space generated by both $b(O)$ and the debiasing covariates induced by the propensity-score model. When the OR model is misspecified, this enlarged projection space can substantially reduce the residual variance, yielding a strict efficiency gain without changing the basic doubly robust logic. 
This variance-reduction property distinguishes GEC from both TMLE
\citep{vanderlaan2011targeted} and the regularized calibrated estimation of
\citet{tan2020regularized}. 

\begin{remark}
\label{rem:local-eff}
When both models are correctly specified and $b^*(O) = \mathbb{E}\{U(\theta;Z)\mid O\}$ is used in calibration, both $V_1$ and $V_3$ reduce to
\begin{align*}
    V_1^* \;&=\; \mathrm{Var}\bigl\{U(\theta_0;Z)\bigr\} \\
    &\qquad \;+\; \mathbb{E}\!\left[\left(\frac{1}{\pi(O;\varphi_0)} - 1\right) \mathrm{Var}\{U(\theta_0;Z)\mid O\}\right],
\end{align*}
which coincides with the semiparametric efficiency lower bound of \citet{robins1994estimation}. Thus, the proposed estimator is also locally efficient. 
\end{remark}

\section{Computation and Cross-Fitting}\label{sec:computation}

The proposed framework is defined through calibration constraints, but its practical implementation is straightforward. The main computational feature distinguishing the present setting from standard entropy calibration is that the balancing constraint depends on the target parameter $\theta$. As a result, the calibration weights themselves depend on $\theta$, and estimation proceeds through a profile optimization: for a candidate value of $\theta$, we first solve the calibration problem to obtain the corresponding weights, and we then update $\theta$ using the resulting weighted estimating equation. Thus, the method combines a low-dimensional convex calibration step with an outer parameter-update step.

This nested structure is the computational counterpart of the statistical formulation in Section~\ref{sec:proposed}. The inner step enforces the balancing and debiasing constraints through generalized entropy calibration, while the outer step updates the parameter of interest using the calibrated weighted estimating equation. Because the inner problem is convex in the dual variable $\lambda$, each calibration update is computationally stable and can be solved with standard optimization routines.

Algorithm~\ref{alg:twoloop} summarizes the resulting two-loop profile optimization procedure for computing $\hat\theta_\omega$.

\subsection{Profile optimization}\label{sec:profile}

The optimal calibration function is $b^*(\theta; O_i) = \mathbb{E}\{U(\theta; Z_i) \mid O_i\}$, which depends on the unknown parameter~$\theta$.
Under this choice, the dual objective becomes $\rho_G(\lambda, \theta)$, with $s_i(\theta) = (b^{*\top}(\theta; O_i),\, g_i(\hat\phi))^\top$, and the estimator solves the saddle-point problem
\begin{equation}\label{eq:saddle}
  \hat\theta_\omega = \arg\max_\theta \min_\lambda\; \rho_G(\lambda, \theta).
\end{equation}
Equivalently, defining the profile objective $L(\theta) = \rho_G(\hat\lambda(\theta), \theta)$ where $\hat\lambda(\theta) = \arg\min_\lambda \rho_G(\lambda, \theta)$, we have $$\hat\theta_\omega = \arg\max_\theta L(\theta).$$

We solve \eqref{eq:saddle} using a two-loop procedure.
For a fixed~$\theta$, the \emph{inner loop} minimizes $\rho_G(\lambda, \theta)$ over~$\lambda$ using Newton's method.
The \emph{outer loop} updates $\theta$ via a Newton or BFGS step based on the gradient of~$L(\theta)$.
By the envelope theorem \citep{danskin2012theory}, $\nabla_\theta L(\theta) = \nabla_\theta \rho_G(\hat\lambda(\theta), \theta)$, since $\nabla_\lambda \rho_G = 0$ at the inner optimum.
The procedure is summarized in Algorithm~\ref{alg:twoloop}.

\begin{algorithm}
\caption{Two-Loop Profile Optimization for $\hat\theta_\omega$. At a high level, the algorithm alternates between calibrating the weights for the current value of $\theta$ and updating $\theta$ using the resulting profile objective.
}\label{alg:twoloop}
\begin{algorithmic}[1]
\Require Initial value $\theta^{(0)}$; tolerances $\epsilon > 0$, $\delta > 0$
\State $k \gets 0$
\Repeat
  \State \textbf{Inner loop:} Compute $\hat\lambda(\theta^{(k)}) = \arg\min_\lambda \rho_G(\lambda, \theta^{(k)})$
  \State \textbf{Outer loop:} Update \[\theta^{(k+1)} \gets \theta^{(k)} - [\nabla^2_\theta L(\theta^{(k)})]^{-1}\, \nabla_\theta L(\theta^{(k)})\]
  \State $k \gets k + 1$
\Until{$\|\theta^{(k)} - \theta^{(k-1)}\| \leq \epsilon$ or $\|\nabla_\theta L(\theta^{(k-1)})\| \leq \delta$}
\State \Return $\hat\theta_\omega \gets \theta^{(k)}$
\end{algorithmic}
\end{algorithm}

\subsection{Calibration via cross-fitting}\label{sec:crossfit}

In practice, the ideal calibration function
\[
b^*(\theta;O_i)=\mathbb{E}\{U(\theta;Z_i)\mid O_i\}
\]
is unknown and must be approximated from the data. This is the point at which modern prediction enters the framework. Rather than specifying a parametric outcome-regression model, we approximate $b^*(\theta;O_i)$ by predicting the missing component $M_i$ from the observed component $O_i$ and then plugging that prediction into the estimating function:
\[
b^*(\theta;O_i)\approx U(\theta;O_i,\hat M_i).
\]
This yields a prediction-powered version of generalized entropy calibration in which the balancing constraint is constructed from data-adaptive imputations of the missing part of the full-data vector.

To avoid the overfitting bias that would arise if the same observations were used both to train the prediction rule and to define the calibration function, we adopt $K$-fold cross-fitting, following \citet{chernozhukov2018double} and \citet{angelopoulos2023prediction}. Cross-fitting ensures that each prediction $\hat M_i$ is obtained from a model trained on data that exclude unit $i$, thereby preserving the separation between the learning step and the estimating step.

Let $\mathcal{I} = \{1, \ldots, N\}$ and $\mathcal{S} = \{i \in \mathcal{I} : \delta_i = 1\}$.
We randomly partition $\mathcal{I}$ into $K$ disjoint folds $\{\mathcal{I}^{(k)}\}_{k=1}^K$.
For each fold~$k$, we fit a prediction model $\hat m^{(-k)}(\cdot)$ using training data $\mathcal{S}^{(-k)} = \mathcal{S} \setminus (\mathcal{S} \cap \mathcal{I}^{(k)})$ and compute out-of-fold predictions $\hat M_i^{(-k)} = \hat m^{(-k)}(O_i)$ for $i \in \mathcal{I}^{(k)}$.

Using these out-of-fold predictions, we evaluate the calibration function for every unit in the sample, including units with $\delta_i=0$. This is essential because the balancing constraint is imposed at the full-sample level: although the weights are attached only to observed units, the calibration targets themselves must be computed for all units.

Using these cross-fitted predictions, we solve the calibration problem:
\begin{align}
&\min_{\omega_1, \ldots, \omega_N} \sum_{i=1}^{N} \delta_i\, G(\omega_i), \label{eq:cf_primal}\\
&\text{subject to}\quad 
\sum_{k=1}^{K}\sum_{i \in \mathcal{I}^{(k)}} \delta_i\, \omega_i\, 
U\!\left(\theta;\, O_i, \hat M_i^{(-k)}\right) \nonumber\\
&\hspace{3.5cm} = 
\sum_{k=1}^{K}\sum_{i \in \mathcal{I}^{(k)}} 
U\!\left(\theta;\, O_i, \hat M_i^{(-k)}\right), \label{eq:cf_balance}\\
&\phantom{\text{subject to}}\quad 
\sum_{i=1}^{N} \delta_i\, \omega_i\, g(\hat\pi_i^{-1}) 
= \sum_{i=1}^{N} g(\hat\pi_i^{-1}). \label{eq:cf_debias}
\end{align}
The two-loop procedure in Algorithm~\ref{alg:twoloop} is then applied to obtain $\hat\theta_\omega$.

This cross-fitted formulation preserves the same two-part logic as the original calibration problem. The balancing constraint uses out-of-fold predictions to approximate the conditional mean structure of the estimating function, while the debiasing constraint continues to anchor the weights to the propensity-score model. Thus, cross-fitting does not alter the inferential architecture of the method; it simply replaces a parametric calibration function by a flexible learned approximation.

Note that  the prediction model $\hat m^{(-k)}$ is trained only on the complete cases in $\mathcal{S}^{(-k)}$, but the out-of-fold predictions $\hat M_i^{(-k)}$ are computed for \emph{all} units $i \in \mathcal{I}^{(k)}$, including those with $\delta_i = 0$.
This is essential: the balancing constraint \eqref{eq:cf_balance} involves the full sample, so calibration function values are needed for every unit regardless of whether $M_i$ is observed.

\begin{remark}
Under correct specification of the propensity-score model, the cross-fitted predictor $\hat M^{(-k)}$ needs only to consistently approximate the conditional mean of the missing component in mean square:
\[
E\|\hat M_i^{(-k)}-E(M_i\mid O_i)\|^2=o(1).
\]
No specific rate such as $N^{-1/4}$ is required for first-order validity along this PS-correct pathway, because the debiasing constraint anchors the calibration weights to the IPW solution even when the predictor converges relatively slowly. Stronger conditions may be needed when first-order validity relies instead on the OR pathway; see \cite{leekim2026} for more technical details. 
\end{remark}

\section{Illustrative Examples}\label{sec:applications}

We now illustrate how the proposed framework specializes to three important settings: causal inference, semi-supervised learning, and regression with missing covariates. The goal of this section is not merely to present three applications of the same algorithm, but to show that these problems share a common inferential structure. In each case, the target parameter is defined by an estimating equation involving a full-data vector $Z_i=(O_i^\top,M_i^\top)^\top$, while only $O_i$ is always observed. The proposed method proceeds in the same way across all three settings: we identify the observed part $O_i$, the missing part $M_i$, the estimating function $U(\theta;Z_i)$, and a calibration function $b(\theta;O_i)$ approximating $\mathbb{E}\{U(\theta;Z_i)\mid O_i\}$. The calibration weights and the resulting estimator are then obtained by applying the general procedure in Sections~\ref{sec:proposed}--\ref{sec:computation}.

These examples clarify the unifying role of calibration. In causal inference, the missing component is a counterfactual outcome; in semi-supervised learning, it is an unlabeled response; in regression with missing covariates, it is an incompletely observed predictor. Although the substantive interpretations differ, the same calibration logic applies in all three cases. Table \ref{tab:2} summarizes the missing data framework for the partially observed data settings. 
 
\begin{table}[t]
\label{tab:2}
\centering
\caption{Common partially observed-data template across the three applications.}
\label{tab:applications}
\begin{tabular}{lllll}
\toprule
Setting & $O_i$ & $M_i$ & $b(\theta;O_i)$ \\
\midrule
Causal inference & $X_i$ & $Y_i(t)$  & $\hat Y_i(t)-\theta_t$ \\
Semi-supervised  learning & $X_i$ & $Y_i$ & $U(\theta;X_i,\hat Y_i)$ \\
Missing covariates & $(X_{1i},Y_i)$ & $X_{2i}$  &
$U(\beta;X_{1i},\hat X_{2i},Y_i)$  \\
\bottomrule
\end{tabular}
\end{table}

\subsection{Causal inference}\label{sec:app_causal}

We begin with causal inference, where the missing-data interpretation is especially transparent. For each unit, one potential outcome is observed and the other is missing, so estimation of the average treatment effect can be viewed as inference with a partially observed full-data vector. Let $T_i\in\{0,1\}$ denote treatment assignment, let $Y_i(1)$ and $Y_i(0)$ be the potential outcomes, and let $X_i\in\mathbb{R}^p$ be pretreatment covariates. Under SUTVA, unconfoundedness, and positivity \citep{rubin1974estimating,RosenRubin83,imbens2015causal}, the observed outcome is
\[
Y_i=T_iY_i(1)+(1-T_i)Y_i(0),
\]
and the average treatment effect is
\[
\theta=\theta_1-\theta_0=\E\{Y(1)\}-\E\{Y(0)\}, 
\]
where $\theta_t$ solves $\mathbb{E}\{U_t(\theta_t; X, Y(t))\} = 0$ with $$U_t(\theta_t; X, Y(t)) = Y(t) - \theta_t \,\,\text{for}\,\, t \in \{0,1\}.$$ 

We estimate $\theta_1$ and $\theta_0$ separately.
For each $t \in \{0,1\}$, define $\delta_i = \mathds{1}(T_i = t)$, $O_i = X_i$, $M_i = Y_i(t)$, and $\pi(X_i; \phi) = P(T_i = t \mid X_i)$.
The calibration function is $b(\theta_t; X_i) = U_t(\theta_t; X_i, \hat Y_i(t)) = \hat Y_i(t) - \theta_t$, where $\hat Y_i(t)$ are cross-fitted predictions.

For the treated group ($t = 1$), the calibration weights $\{\omega_{1i}\}_{i=1}^N$ are obtained by solving
\begin{align}
  &\min_{\omega_{11}, \ldots, \omega_{1N}} \sum_{i=1}^{N} T_i\, G(\omega_{1i}), \label{eq:causal_primal}\\[4pt]
  &\text{subject to}\quad \sum_{i=1}^{N} T_i\, \omega_{1i} = N, \label{eq:causal_norm}\\[4pt]
  &\phantom{\text{subject to}}\quad \sum_{i=1}^{N} T_i\, \omega_{1i}\, \bigl(\hat Y_i(1) - \theta_1\bigr) = \sum_{i=1}^{N} \bigl(\hat Y_i(1) - \theta_1\bigr), \label{eq:causal_balance}\\[4pt]
  &\phantom{\text{subject to}}\quad \sum_{i=1}^{N} T_i\, \omega_{1i}\, g(\hat\pi_i^{-1}) = \sum_{i=1}^{N} g(\hat\pi_i^{-1}), \label{eq:causal_debias}
\end{align}
where $\hat\pi_i = P(T_i = 1 \mid X_i; \hat\phi)$ is the estimated propensity score.
An analogous calibration problem is solved within the control group ($t = 0$) to obtain weights $\{\omega_{0i}\}_{i=1}^N$ and the estimate $\hat\theta_0$.
Because the balancing constraint \eqref{eq:causal_balance} depends on $\theta_t$, we apply the two-loop profile optimization (Algorithm~\ref{alg:twoloop}) within each treatment group to jointly estimate $\theta_t$ and the corresponding weights, and form $\hat\theta = \hat\theta_1 - \hat\theta_0$.
Thus, treatment assignment plays the role of a response indicator, and the proposed estimator treats the unobserved potential outcome exactly as a missing component of the full-data vector.

\subsection{Semi-supervised learning}\label{sec:app_ssl}

Semi-supervised learning provides a second example in which the missing-data interpretation is immediate. Here covariates are observed for a large unlabeled sample, but outcomes are available only for a labeled subset. From the perspective of the present framework, the unlabeled units are simply observations with missing responses. The target parameter is again defined through a full-data estimating equation, and the main statistical task is to recover that equation efficiently using both the labeled and unlabeled samples.

Let $X_i$ denote the covariates and $Y_i$ the outcome of interest. Define $\delta_i=1$ if $Y_i$ is observed and $\delta_i=0$ otherwise, so that $O_i=X_i$ and $M_i=Y_i$. Under MAR,
\[
\delta_i \perp Y_i \mid X_i,
\]
the response mechanism can be modeled through $\pi(X_i;\phi)=P(\delta_i=1\mid X_i)$.

Formally, a labeled dataset $\mathcal{L} = \{(x_i, y_i); i = 1, \ldots, n\}$ with fully observed outcomes is supplemented by an unlabeled dataset $\mathcal{U} = \{x_i; i = n+1, \ldots, n+N\}$.
Here $Z_i = (X_i, Y_i)$, $O_i = X_i$, $M_i = Y_i$, and $\delta_i = 1$ if $Y_i$ is observed.
The target parameter $\theta$ solves $\mathbb{E}\{U(\theta; X, Y)\} = 0$, and the calibration function is $b(\theta; X_i) = U(\theta; X_i, \hat Y_i)$, where $\hat Y_i$ are cross-fitted predictions of $Y_i$ based on $X_i$.

The selection mechanism may be either MAR, with $P(\delta_i = 1 \mid X_i) = \pi(X_i; \phi)$ estimated by logistic regression, or MCAR, with $P(\delta_i = 1) = n/(n+N)$.
In both cases, the calibration weights are obtained from \eqref{eq:cf_primal}--\eqref{eq:cf_debias} and the estimator from Algorithm~\ref{alg:twoloop}.
Concretely, the proposed estimator of $\theta$ solves the weighted estimating equation
\begin{equation}\label{eq:ssl_ee}
  \frac{1}{n+N}\sum_{i=1}^{n+N} \delta_i\, \omega_i\, U(\theta; x_i, y_i) = 0,
\end{equation}
where the weights $\omega_i$ integrate labeled and unlabeled data through the balancing constraint, ensuring that the weighted labeled sample reproduces the covariate moments of the full sample.
This yields a doubly robust semi-supervised estimator that remains efficient even under OR model misspecification: when the prediction model for $Y_i$ is poor, the debiasing constraint based on $\pi(X_i; \hat\phi)$ still corrects for the distributional difference between labeled and unlabeled samples. In this setting, the balancing constraint uses predictions of the unlabeled outcomes to borrow information from the full covariate distribution, while the debiasing constraint corrects for differences between labeled and unlabeled samples when the response mechanism is correctly modeled.

\subsection{Missing covariates in regression}\label{sec:app_mcov}

Our third example concerns regression with missing covariates. Unlike the previous two settings, the outcome is fully observed, while part of the predictor vector is missing. This changes the form of the propensity-score model: because the response indicator refers to covariate observation rather than outcome observation, the probability of observing the missing covariate may depend on the observed response as well as on the fully observed covariates.

Suppose $(Y_i,X_i)$ satisfies the regression model
\[
Y_i=\mu(X_{1i},X_{2i};\beta)+\varepsilon_i,
\qquad
\E(\varepsilon_i\mid X_i)=0,
\]
where $X_i=(X_{1i},X_{2i})$ and $X_{2i}$ is subject to missingness. Here
\[
O_i=(X_{1i},Y_i), \qquad M_i=X_{2i},
\]
and under MAR,
\[
\delta_i \perp X_{2i}\mid (X_{1i},Y_i).
\]
Thus,  the probability of observing $X_{2i}$ may depend on both the fully observed covariates $X_{1i}$ and the response $Y_i$, but not on the missing covariate $X_{2i}$ itself.
This conditioning on~$Y_i$ distinguishes the missing-covariate setting from the missing-outcome case: the propensity score model is
\[
  P(\delta_i = 1 \mid X_{1i}, Y_i) = \pi(X_{1i}, Y_i;\, \phi),
\]
which can be estimated by maximum likelihood, e.g.,
\[
\hat\phi = \arg\max_\phi \sum_{i=1}^{N}
\left[\delta_i \log \pi_i + (1-\delta_i)\log(1-\pi_i)\right],
\]
where $\pi_ i = \pi(X_{1i}, Y_i;\, \phi).$

The estimating function is
\[
  U(\beta;\, X_{1i}, X_{2i}, Y_i) = \{Y_i - \mu(X_{1i}, X_{2i}; \beta)\}\,\frac{\partial}{\partial \beta}\mu(X_{1i}, X_{2i}; \beta),
\]
and the calibration function is $b(\beta; X_{1i}, Y_i) = U(\beta; X_{1i}, \hat X_{2i}, Y_i)$, where $\hat X_{2i}$ are cross-fitted predictions of $X_{2i}$ based on $(X_{1i}, Y_i)$.
The calibration weights and estimator of $\beta$ are obtained by applying \eqref{eq:cf_primal}--\eqref{eq:cf_debias} and Algorithm~\ref{alg:twoloop}.

\section{Simulation Study}\label{sec:simulation}

This section evaluates the finite-sample behavior of the proposed generalized entropy calibration estimators in the three application settings introduced in Section~\ref{sec:applications}. The goal is not simply to compare a collection of competing methods, but to examine whether the empirical results reflect the main theoretical properties established in Section~\ref{sec:theory}. In particular, the simulations are designed to assess four questions: whether the proposed estimators recover the efficient benchmark when both working models are correctly specified; whether they remain accurate when only one of the propensity-score or outcome-regression models is correct; whether they improve upon classical AIPW under correct PS specification and misspecified OR models, as predicted by Corollary~\ref{cor:var-dom}; and whether these qualitative patterns persist across causal inference, semi-supervised learning, and regression with missing covariates.

We report results for the Hellinger distance (HD) and exponential tilting (ET) versions of generalized entropy calibration. Unless otherwise noted, all simulations use $K=4$ folds for cross-fitting.

\subsection{Causal inference}\label{sec:sim_causal}

We begin with causal inference because it provides the clearest setting for isolating the doubly robust behavior of the proposed estimator. The simulation uses a $2\times 2$ factorial design combining two outcome-regression specifications and two propensity-score specifications. This design creates the four regimes most relevant to the theory: joint correctness, OR-only correctness, PS-only correctness, and joint misspecification. It therefore allows us to examine not only double robustness, but also the comparative efficiency gains predicted by Corollary~\ref{cor:var-dom} when the PS model is correct and the OR model is misspecified.

\medskip
\noindent\textbf{Design.}
We consider two outcome-regression models (OR1, OR2) and two propensity-score models (PS1, PS2), using $M=500$ Monte Carlo replicates with sample size $N=1{,}000$. Covariates are generated as $x_i=(x_{i1},x_{i2},x_{i3},x_{i4})^\top\sim N(0_4,I_4)$, and potential outcomes are generated as
\[
y_i(t)=m_t(x_i)+\epsilon_{it},\qquad \epsilon_{it}\sim N(0,1),\quad t\in\{0,1\}.
\]
\begin{itemize}
\item \emph{OR1} (linear): $m_1(x) = 1 + \mathbf{1}^\top \boldsymbol x$, $m_0(x) = \mathbf{1}^\top \boldsymbol x, \mathbf{1} = (1, 1, 1, 1)^\top$; true ATE $= 1$.
\item \emph{OR2} (nonlinear): Let $h(x) = (x-1)^3 - x^2 + x/\{1 + \exp(\mathrm{clip}(x))\} + 10$ with $$\mathrm{clip}(x) = \max\{\min(x,3), -3\}$$ applied componentwise. Set $m_1(x) = 10 + \mathbf{1}^\top\boldsymbol x + \tfrac{1}{2}\mathbf{1}^\top h(\boldsymbol x)$, $m_0(\boldsymbol x) = \mathbf{1}^\top \boldsymbol x + \tfrac{1}{2}\mathbf{1}^\top h(\boldsymbol x)$; true ATE $= 10$.
\item \emph{PS1} (logistic linear): $\pi(\boldsymbol x) = \mathrm{expit}(-0.25 + x_1 + 0.5x_2 - 0.5x_3 - 0.1x_4)$.
\item \emph{PS2} (logistic nonlinear): $\pi(\boldsymbol x) = \mathrm{expit}(x_1 - 0.5x_1 x_2 - x_3^2 + 0.5x_4^3)$.
\end{itemize}
The resulting four combinations of OR and PS models allow us to check whether the proposed calibration estimators match AIPW under joint correctness, remain reliable when only one working model is correct, and exhibit the clearest gains precisely in the PS-correct/OR-misspecified regime emphasized by the theory.

We compare HD and ET with seven existing methods: inverse probability weighting \citep[IPW;][]{horvitz1952}, entropy balancing \citep[EBPS;][]{hainmueller2012}, optimal covariate balancing \citep[oCBPS;][]{fan2022optimal}, covariate balancing propensity score \citep[CBPS;][]{imai2014}, empirical balancing calibration weighting \citep[EBCW;][]{chan2016}, and AIPW with either a linear model or GAM for the outcome regression, denoted AIPW\,(LM) and AIPW\,(GAM).

\begin{figure*}
    \centering
    \includegraphics[scale=.5]{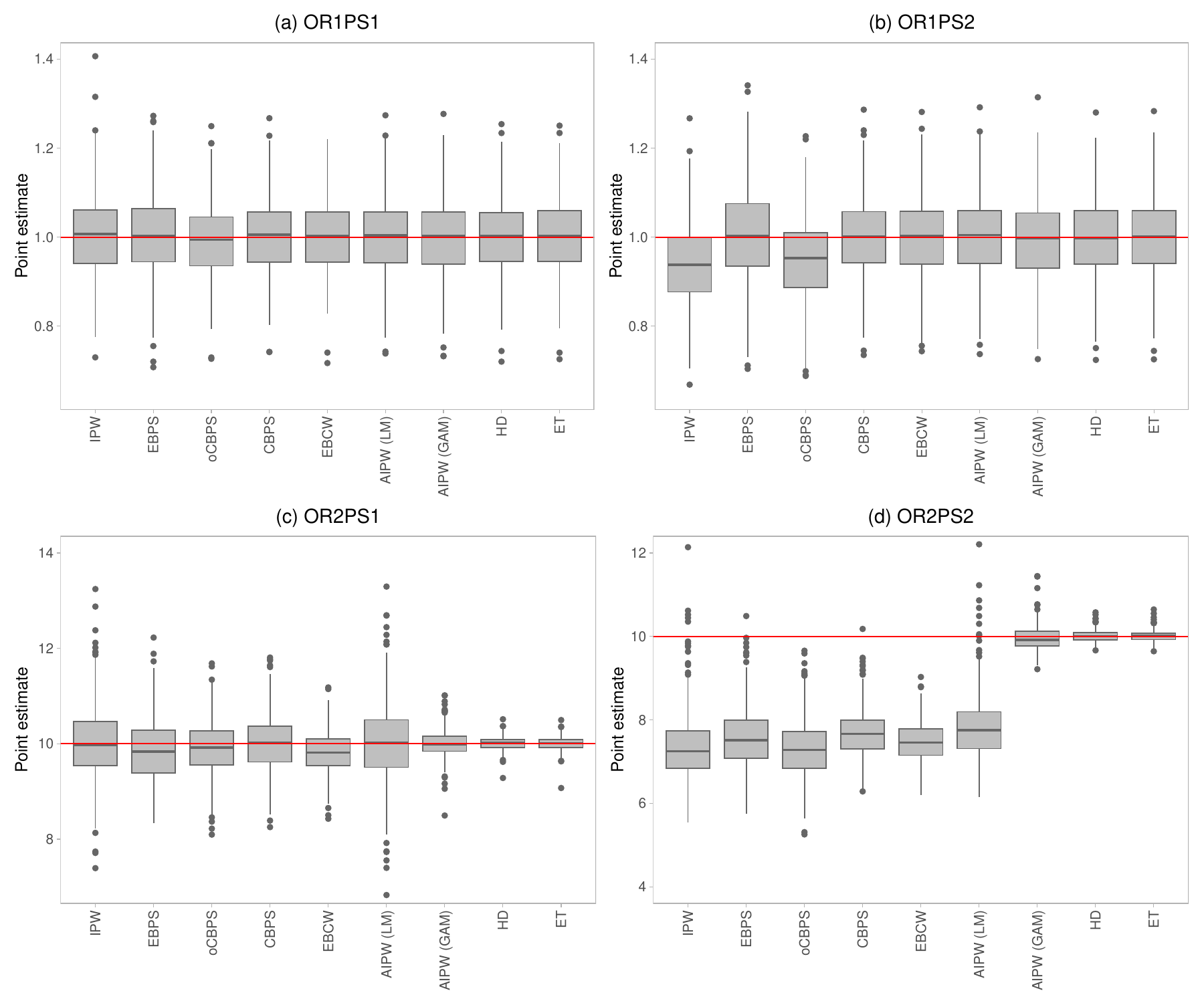}
\caption{Estimation of Average Treatment Effects (ATE) under four scenarios: under
OR1(2), the outcome regression (OR) model is correctly specified (misspecified); under
PS1(2), the propensity score model is correctly specified (misspecified). The horizontal red line represents the true ATE.} 
    \label{fig:ATE}
\end{figure*}

\medskip
\noindent\textbf{Results.}
Figure~\ref{fig:ATE} summarizes the Monte Carlo distributions across four specification regimes.
Under OR1PS1 (both models correct), all methods show negligible bias and similar variability, consistent with the local efficiency result in Remark~\ref{rem:local-eff}: when both models are correctly specified, the proposed estimator matches AIPW.

Under OR1PS2 (PS misspecified, OR correct), IPW and oCBPS exhibit substantial bias because they rely exclusively on the PS model, whereas HD, ET, and the remaining calibration-based and augmented estimators maintain low bias, confirming the OR-based consistency pathway in Corollary~\ref{cor:or}.

The most informative comparison is OR2PS1 (OR misspecified, PS correct), which isolates the variance-reduction mechanism identified in Corollary~\ref{cor:var-dom}.
Here, HD and ET display noticeably smaller interquartile ranges than all competitors, including both AIPW variants. 
Among the AIPW estimators, AIPW\,(GAM) shows a clear efficiency gain over AIPW\,(LM), as the more flexible outcome model partially compensates for misspecification; nevertheless, both AIPW versions remain less efficient than HD and ET.
The weighting-only methods (IPW, EBPS, oCBPS, CBPS, EBCW) show substantially wider boxplots, reflecting their lack of outcome-model augmentation.

Under joint misspecification (OR2PS2), all methods are potentially inconsistent.
Nevertheless, HD and ET remain the most robust, exhibiting the smallest bias and most concentrated sampling distributions, which suggests that the entropy calibration framework degrades gracefully when neither model is fully correct. This is somewhat related to the global robustness of the HD  discussed by \cite{antoine2021}. 

\subsection{Semi-supervised learning}\label{sec:sim_ssl}

We next consider semi-supervised learning to examine whether the same qualitative pattern carries over to a setting in which the missing component is an unlabeled response rather than a counterfactual outcome. From the perspective of the present framework, this is again a partially observed-data problem: covariates are observed for all units, while responses are available only for a labeled subset. The simulation is designed to assess how effectively the proposed calibration estimators use the unlabeled covariates for efficiency, while the debiasing constraint protects against bias when the working prediction model is imperfect or when the labeled and unlabeled samples differ in distribution.

The main question in this experiment is whether the same advantages seen in the causal-inference setting persist when the balancing constraint is constructed from cross-fitted predictions of the missing response. In particular, we examine whether ET and HD continue to recover the efficient benchmark under favorable specification, remain stable under partial misspecification, and improve upon purely supervised or outcome-model-based semi-supervised competitors when the prediction model is imperfect.

\medskip
\noindent\textbf{Design.}
The target is the regression coefficient $\beta$ in $y_i = x_i^\top \beta + \epsilon_i$.
Covariates are $x_i \sim N(0_p, I_p)$ with $p = 4$, total sample size $n + N = 2{,}000$, and $M = 1{,}000$ replications.
The outcome is generated under two specifications:
\begin{itemize}
\item \emph{OR1} (linear): $y_i = \alpha_0 + \alpha_1^\top x_i + \epsilon_i$;
\item \emph{OR2} (nonlinear): $y_i = \alpha_0 + \alpha_1^\top x_i + \alpha_2^\top\{x_i^3 - x_i^2 + \exp(x_i)\} + \epsilon_i$,
\end{itemize}
where $(\alpha_0, \alpha_1^\top, \alpha_2^\top)^\top = (1, \mathbf{1}^\top, \mathbf{1}^\top)^\top$ and $\epsilon_i \sim N(0,1)$.
The true value of the target parameter $\beta$ is computed by generating data of size $10^7$.
Note that under OR2 the data-generating model is nonlinear, so the linear working model $y_i = x_i^\top \beta + \epsilon_i$ is misspecified; nevertheless, $\beta$ remains a well-defined projection parameter that our method targets.
Two selection mechanisms are considered:
\emph{MAR}, with $\mathrm{logit}(p_i) = -1 - x_{1i} - 0.5 x_{2i} + 0.5 x_{3i} + 0.1 x_{4i}$;
and \emph{MCAR}, with $p_i = n/(n+N)$.
We compare HD and ET with the supervised estimator (Sup) and four semi-supervised methods: the projection-based estimator \citep[PSSE;][]{song2023general}, density-ratio estimation \citep[DRESS;][]{kawakita2013semi}, efficient adaptive estimation \citep[EASE;][]{chakrabortty2018efficient}, and the partial-information estimator \citep[PI;][]{azriel2022semi}.

\begin{figure*}
\includegraphics[scale=.45]{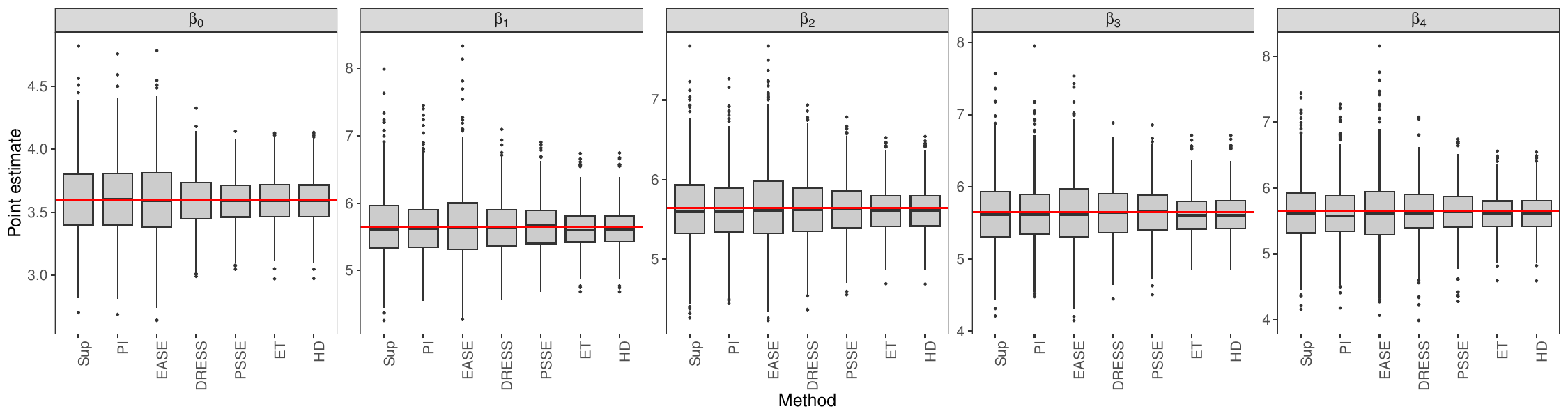}
    \caption{Boxplots of the estimated linear regression coefficients $\beta$ under OR2 (OR model misspecification) with MAR missingness. }
    \label{fig:MAR_OR2PS1}
\end{figure*}

%Figures \ref{fig:MAR_OR2PS1} and \ref{fig:MCAR_OR2PS1}
\medskip
\noindent\textbf{Results.}
Under OR model misspecification (OR2) with MAR (Figure~\ref{fig:MAR_OR2PS1}), HD and ET achieve the highest efficiency among all methods for all five regression coefficients $\beta_0, \ldots, \beta_4$.
The improvement is most visible for $\beta_0$: the interquartile range of HD and ET is roughly half that of the supervised estimator, and noticeably smaller than those of PI, EASE, DRESS, and PSSE.
The existing semi-supervised methods (PI, EASE, DRESS, PSSE) provide only modest gains over the supervised estimator under OR misspecification, because their efficiency improvements rely on the outcome model being approximately correct.
In contrast, our proposed estimators benefit from the debiasing constraint, which compensates for outcome-model misspecification through the propensity-score information.

\begin{figure*}
\includegraphics[scale=.45]{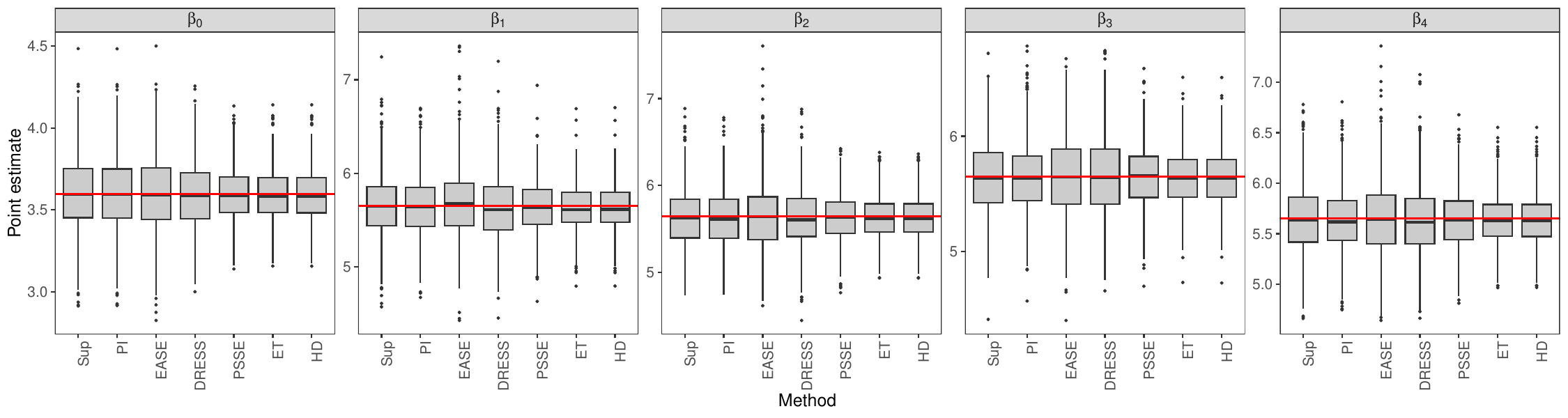}
    \caption{Boxplots of the estimated linear regression coefficients $\beta$ under OR2 (OR model misspecification) with MCAR missingness. }
    \label{fig:MCAR_OR2PS1}
\end{figure*}

Under OR2 with MCAR (Figure~\ref{fig:MCAR_OR2PS1}), the same pattern holds: HD and ET dominate in efficiency.
The MCAR mechanism simplifies the selection model, so the debiasing constraint is particularly effective, and the efficiency gains of HD and ET over competitors are even more pronounced.

\begin{figure*}
\includegraphics[scale=.45]{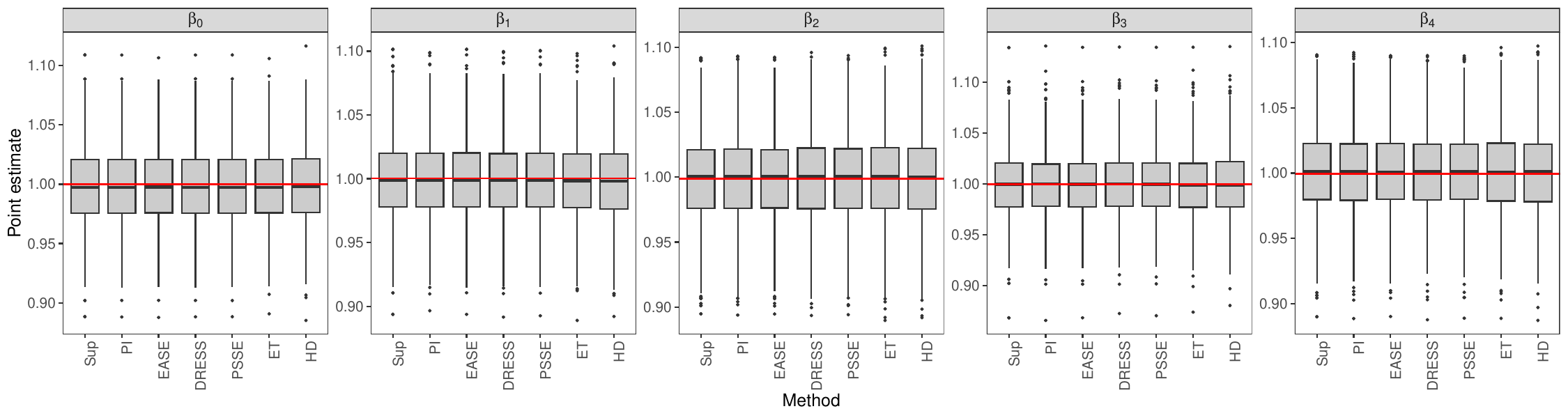}
    \caption{Estimation of the linear regression coefficients $\beta$  under OR1 (when the OR model is correctly specified) and MCAR missingness mechanism. }
    \label{fig:MCAR_OR1PS1}
\end{figure*}

Under a correctly specified OR model (OR1) with MCAR (Figure~\ref{fig:MCAR_OR1PS1}), all semi-supervised estimators---including HD, ET, PI, EASE, DRESS, and PSSE---are consistent and nearly as efficient as the supervised estimator, confirming that the proposed method does not sacrifice efficiency when the outcome model is correct.
This is consistent with the local efficiency result (Remark~\ref{rem:local-eff}): when both models are correctly specified, the calibration weights are close to unity and the estimator behaves like the full-sample supervised estimator.

\subsection{Missing covariates in regression}\label{sec:sim_misscov}

Our third simulation considers regression with missing covariates. This setting is conceptually important because the incompletely observed component is now a predictor rather than an outcome, and the response mechanism may depend on the observed response as well as on the fully observed covariates. The purpose of this experiment is to determine whether the same calibration logic continues to deliver the expected robustness and efficiency gains in a structurally different missing-data problem.

As in the previous settings, the comparison is organized around the behavior of the method under correct and misspecified working models. Of particular interest is the regime in which the propensity-score model is correctly specified but the regression-based imputation model for the missing covariate is misspecified. According to Corollary~\ref{cor:var-dom}, this is the setting in which the debiasing constraint should provide the clearest improvement over classical AIPW by enlarging the effective calibration space. The simulation therefore serves as a direct finite-sample check of whether the variance-dominance phenomenon identified in Section~\ref{sec:theory} extends beyond missing outcomes to the case of incompletely observed predictors.

\medskip
\noindent\textbf{Design.}
We estimate $\beta = (\beta_0, \beta_1, \beta_2)$ in the linear model $y_i = \beta_0 + \beta_1 x_{1i} + \beta_2 x_{2i} + \epsilon_i$, where $x_{1i} \sim N(0,1)$, $x_{2i} \sim \mathrm{Bernoulli}(0.5)$, $\epsilon_i \sim N(0,1)$, and $x_{2i}$ is subject to missingness.
Sample size is $N = 500$ with $M = 1{,}000$ replications.
Two PS models govern the missingness of $x_{2i}$:
\emph{PS1}~(MAR), $\mathrm{logit}(\pi_i) = -1 + 0.5 x_{1i} + 0.5 y_i$;
\emph{PS2}~(MCAR), $\mathrm{logit}(\pi_i) = -1$.
Two OR models generate $y_i$:
\begin{itemize}
\item \emph{OR1}: $y_i = 1 + x_{1i} + 2x_{2i} + \epsilon_{1i}$, $\epsilon_{1i} \sim N(0,1)$;
\item \emph{OR2}: $y_i = 0.5 + 2\sin(\pi x_{1i}) - 1.5\cos(2\pi x_{1i}) + 0.25 x_{1i}^3 + x_{2i} + \epsilon_{2i}$, $\epsilon_{2i} \sim N(0,4)$.
\end{itemize}
Cross-fitted predictions $\hat x_{2i}$ are obtained via logistic regression, since $x_{2i}$ is binary.
We compare the proposed HD estimator with four benchmarks: the full-sample estimator (Full, infeasible), the complete-case estimator (CC), the Horvitz--Thompson estimator (HT), and the AIPW estimator.

\medskip
\noindent\textbf{Results.}
Table~\ref{tab:covariate_missing} reports bias, standard error, and RMSE ($\times 10$) across all scenarios.
The full-sample estimator serves as a benchmark with negligible bias and smallest RMSE.
The CC estimator shows substantial bias under MAR (OR1PS1 and OR2PS1), confirming that discarding incomplete cases introduces systematic error when the missingness depends on the observed data.

Under OR1PS1 (both models correct), HD achieves RMSE comparable to or slightly better than AIPW for all three coefficients (e.g., RMSE of $0.71$ vs.\ $0.75$ for $\beta_0$, and $1.12$ vs.\ $1.18$ for $\beta_2$).
Both HT and AIPW reduce bias relative to CC but exhibit inflated standard errors.

The largest efficiency gain appears under OR2PS1 (OR misspecified, PS correct).
Here, HD achieves RMSE of $1.31$, $0.92$, and $2.48$ for $\beta_0$, $\beta_1$, and $\beta_2$ respectively, compared with $1.93$, $1.44$, and $3.79$ for AIPW---a reduction of approximately $32\%$, $36\%$, and $35\%$.
The HT estimator, which does not use outcome information, performs comparably to AIPW or worse, highlighting the value of incorporating the calibration function even when the outcome model is imperfect.

Under MCAR (PS2), all weighting methods perform comparably because the selection mechanism is simple and correctly specified by all methods, with HD showing a slight advantage in OR1PS2 (e.g., RMSE of $0.87$ vs.\ $0.90$ for $\beta_2$) and essentially equivalent performance in OR2PS2.

Across all three settings, the simulations tell a consistent story. When both working models are correct, the proposed estimators recover the efficient doubly robust benchmark. When only one working model is correct, they retain the expected robustness. Most importantly, when the PS model is correct and the OR model is misspecified, ET and HD often improve upon classical AIPW, which is the finite-sample pattern predicted by Corollary~\ref{cor:var-dom}.

\begin{table*}
    \centering
    \small
    \caption{ Bias ($\times 10$), standard error (SE) ($\times 10$) and root mean square error (RMSE) ($\times 10$) for estimators under missing covariates in regression model setting.}
    
    \begin{tabular}{cccccccccccccccc}
    
        \toprule
       Model &Method & \multicolumn{3}{c}{$\beta_0=1$} & \multicolumn{3}{c}{$\beta_1=1$}  & \multicolumn{3}{c}{$\beta_2=2$}\\
        \cmidrule(lr){3-5} 
        \cmidrule(lr){6-8} \cmidrule(lr){9-11}
       & & Bias & SE & RMSE & Bias & SE & RMSE & Bias & SE & RMSE\\
        \midrule
         &Full  & -0.01 & 0.43 & 0.43 & -0.03 & 0.33 & 0.33 &  0.04 & 0.63 & 0.64 \\
&CC    &  3.83 & 1.00 & 3.96 & -1.01 & 0.63 & 1.19 & -0.94 & 1.16 & 1.50 \\
OR1PS1&HT    &  0.14 & 1.38 & 1.39 & -0.13 & 1.01 & 1.02 & -0.09 & 1.79 & 1.79 \\
&AIPW  & -0.07 & 0.75 & 0.75 & -0.03 & 0.82 & 0.82 &  0.16 & 1.17 & 1.18 \\
&HD& -0.12 & 0.70 & 0.71 & -0.08 & 0.74 & 0.75 &  0.35 & 1.07 & 1.12 \\
      \hline 
      & Full  & -0.01 & 0.43 & 0.43 & -0.03 & 0.33 & 0.33 &  0.04 & 0.63 & 0.64 \\
& CC    &  0.01 & 0.85 & 0.85 & -0.04 & 0.62 & 0.62 &  0.05 & 1.24 & 1.24 \\
OR1PS2& HT    &  0.00 & 0.77 & 0.77 & -0.04 & 0.62 & 0.62 &  0.05 & 1.24 & 1.24 \\
& AIPW  & -0.02 & 0.64 & 0.64 & -0.03 & 0.50 & 0.50 &  0.09 & 0.90 & 0.90 \\
& HD    & -0.04 & 0.63 & 0.63 & -0.03 & 0.50 & 0.50 &  0.13 & 0.86 & 0.87 \\
\hline
& Full  &  0.00 & 0.94 & 0.94 &  0.00 & 0.85 & 0.85 &  0.01 & 1.30 & 1.30 \\
& CC    & 11.44 & 1.56 & 11.54 & -2.66 & 1.44 & 3.02 & -2.05 & 2.00 & 2.86 \\
OR2PS1& HT    &  0.51 & 1.88 & 1.95 & -0.98 & 2.33 & 2.53 & -0.15 & 2.77 & 2.77 \\
& AIPW  & -0.03 & 1.93 & 1.93 & -0.09 & 1.43 & 1.44 &  0.21 & 3.78 & 3.79 \\
& HD    & -0.09 & 1.31 & 1.31 &  0.00 & 0.92 & 0.92 &  0.35 & 2.45 & 2.48 \\
\hline
& Full  & -0.02 & 1.22 & 1.22 & -0.03 & 1.02 & 1.02 & 0.04 & 1.72 & 1.72 \\
& CC    &  0.00 & 2.20 & 2.20 & -0.13 & 2.03 & 2.04 & 0.05 & 3.30 & 3.30 \\
OR2PS2& HT    & -0.01 & 1.81 & 1.81 & -0.14 & 2.02 & 2.03 & 0.05 & 3.32 & 3.32 \\
& AIPW  & -0.02 & 1.78 & 1.78 & -0.02 & 1.07 & 1.07 & 0.08 & 3.26 & 3.26 \\
& HD    & -0.08 & 1.84 & 1.84 & -0.02 & 1.08 & 1.08 & 0.20 & 3.36 & 3.37 \\
\hline
    \end{tabular}
    
    \label{tab:covariate_missing}
\end{table*}

\section{Real-Data Applications}\label{sec:realdata}

We conclude with two real-data applications illustrating the practical scope of the proposed framework. The goal of this section is not merely to compare numerical performance across methods, but to show how the same calibration logic applies in substantively different partially observed-data problems. The first application is a causal-inference benchmark, where one may assess performance against an experimental reference and examine covariate balance directly. The second is a semi-supervised learning problem from public health, where outcomes are observed only for a labeled subset and efficiency gains from unlabeled covariates are of primary interest.

Taken together, these examples illustrate two complementary aspects of the framework. In the causal setting, calibration helps reconstruct the target treatment comparison from observational controls while maintaining close balance on relevant covariates. In the semi-supervised setting, calibration uses predictions for the unlabeled outcomes to improve efficiency while preserving a debiasing correction tied to the observation mechanism.

\subsection{Causal inference: the LaLonde job-training study}\label{sec:real_causal}

We first consider the National Supported Work (NSW) Demonstration, a canonical benchmark for causal inference with observational controls \citep{lalonde1986evaluating, dehejia1999causal}. This application is useful for two reasons. First, the experimental NSW estimate provides a reference against which observational estimators can be compared through evaluation bias. Second, the weighting structure can be examined directly through covariate balance, making the role of calibration transparent.

Following the standard benchmark design, we pool the NSW experimental sample with two large observational control groups, the Panel Study of Income Dynamics (PSID) and the Current Population Survey (CPS), into a single sample. Units are classified by a group indicator $G_i\in\{1,2,3,4\}$: NSW treated ($N_1=185$), NSW controls ($N_2=260$), PSID ($N_3=2{,}490$), and CPS ($N_4=15{,}992$). The target estimand is the average treatment effect for the NSW experimental population,
\[
ATE_{\text{NSW}} \equiv E\{Y(1)-Y(0)\mid \text{NSW participants}\}.
\]
We estimate generalized propensity scores $\pi_{ig}=P(G_i=g\mid x_i)$ via multinomial logistic regression and define tilting weights $r_{ig}=\pi_{i,\text{NSW}}/\pi_{ig}$, where $\pi_{i,\text{NSW}}=\pi_{i1}+\pi_{i2}$. The covariates used for propensity-score estimation are age, years of education, indicators for Black and Hispanic ethnicity, marital status, an indicator for no high school degree, and real earnings in 1974 and 1975 (re74 and re75), following the standard LaLonde benchmark literature. Cross-fitted outcome predictions $\hat y_i(g)$ are obtained using GAM with $K=4$ folds. For each candidate control group $g\in\{2,3,4\}$, calibration weights are constructed by solving (18)--(20) with the two-loop procedure, and the ATE is estimated as $\widehat{ATE}^{[g]}=\bar Y_1-\hat\theta_g$, where $\bar Y_1$ is the mean outcome among NSW treated units.

\noindent\textbf{Results.}
Table~\ref{tab:nsw} shows that generalized entropy calibration performs competitively with established weighting and doubly robust methods in both external-control settings, while remaining close to the experimental benchmark $\hat\theta_{\text{NSW}}=1794$. Using the NSW experimental data alone, all estimators are close to the benchmark, as expected under randomization. Using PSID or CPS as external controls, the unweighted difference in means exhibits very large evaluation bias, reflecting substantial covariate imbalance between the NSW treated group and the observational control samples. Among the weighting estimators, ET yields one of the smallest evaluation biases in both settings: for PSID, its absolute bias is smaller than that of IPW, CBPS, EBCW, and the two AIPW implementations; for CPS, its estimate is again close to the benchmark and broadly comparable to the strongest competing approaches, with a standard error similar to or smaller than several alternatives. Overall, the pattern suggests that generalized entropy calibration provides a favorable balance between bias reduction and precision across both external-control samples. 

\begin{figure*}
    \centering
    \includegraphics[scale=.5]{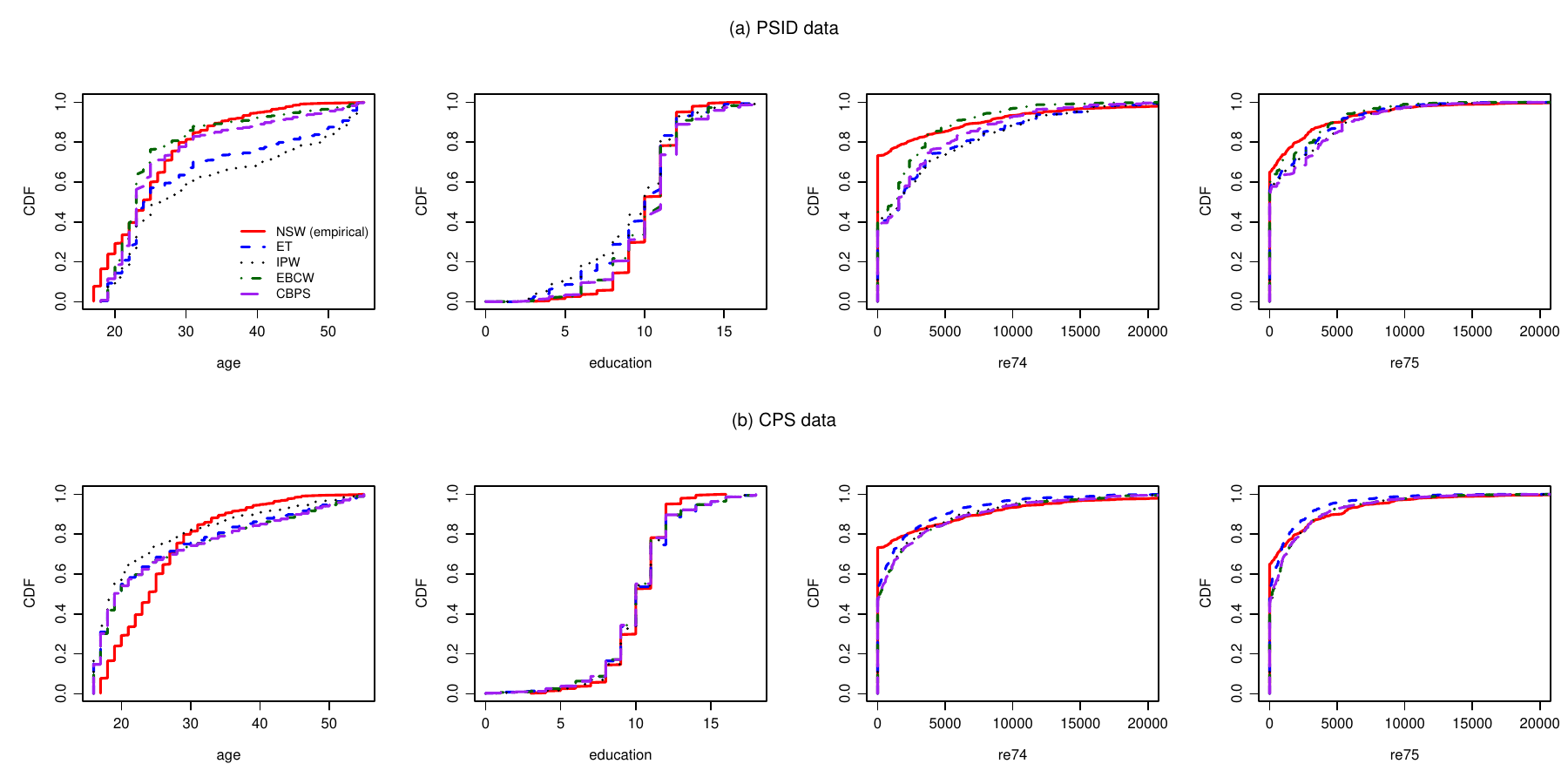}
\caption{Weighted covariate distributions for the \cite{lalonde1986evaluating} data.} 
    \label{fig:balance}
\end{figure*}

Figure~\ref{fig:balance} compares weighted covariate distributions for age, education, re74, and re75 in the PSID and CPS control samples against the empirical distribution of the combined NSW sample. The figure shows that generalized entropy calibration achieves close alignment across these key variables in both panels. For the PSID controls, the contrast is especially visible: IPW retains noticeable residual imbalance in age and re74, while CBPS and EBCW improve the fit but still leave discrepancies in parts of the earnings distributions. For the CPS controls, all methods perform more similarly because the untreated sample is already closer to the NSW sample, though ET and EBCW continue to show the closest overall agreement. These balance patterns are consistent with the evaluation-bias comparisons in Table~\ref{tab:nsw}: methods that achieve better covariate balance tend to exhibit smaller evaluation bias. This application therefore illustrates the causal-inference face of the proposed framework: calibration improves agreement with the experimental benchmark not simply by reweighting, but by enforcing a balance structure that better aligns the observational controls with the treated sample.

\begin{table*}
\setlength{\tabcolsep}{5pt}
\caption{Comparison of Average Treatment Effect Estimators on the NSW Experimental and PSID and CPS observational data. \\ Notes:  SE= standard error, EB = Evaluation Bias. }

\begin{tabular}{l
  c@{\hspace{5pt}}c @{\hspace{15pt}}
  c@{\hspace{5pt}}c @{\hspace{10pt}}
  c@{\hspace{15pt}}c@{\hspace{2pt}}
  c c c c c c c}
\hline
\multicolumn{11}{c}{\textbf{(a) Treatment effects}} \\
\midrule
& \multicolumn{2}{c}{\textbf{NSW Data}}
& \multicolumn{3}{c}{\textbf{PSID Data}}
& \multicolumn{3}{c}{\textbf{CPS Data}}
\\
\cmidrule(lr){2-3}\cmidrule(lr){4-6}\cmidrule(lr){7-9}
\textbf{Estimators}
& \textbf{Estimates} & \textbf{SE}
& \textbf{Estimates} & \textbf{EB}& \textbf{SE}
& \textbf{Estimates} & \textbf{EB}& \textbf{SE}
&  &  &  & & \\
\midrule
Unweighted         & 1794 & 671&   -15205 & -17000&    657&   -8498  &-10292& 583 \\
IPW                            & 1796 & 673&1474 &-320&901&1064 &-730& 644 \\

CBPS &  1636 & 687&1389 &-405& 886&1288&  -506&641\\
EBCW                           & 1792&666 &2316  &522&799 &1284 &-510 &633 \\
%AIPW&1786  &  667&1436 &-358 & 876 &1684& -110  &650\\

AIPW (LM)    & 1779 & 673 &  1283 &-511& 909&1291& -503 & 652\\
AIPW (GAM)    & 1780 & 672 &  1426 &-368& 893&1683& -111 & 656\\
%HD                             & 1651 & 695 &2655 & 861 & 865&1012 &-782 &1026 \\
%ET                            & 1783  & 666 & 1589 &-205&  923& 1626 &-168 & 634\\
ET                            & 1787 & 670 & 1655 &-139& 912& 1615 & -179&635\\

\bottomrule
\end{tabular}

\label{tab:nsw}
\end{table*}

\subsection{Semi-supervised learning: NHANES fasting glucose}\label{sec:real_ssl}

We next consider a semi-supervised regression problem using the 2017--2018 National Health and Nutrition Examination Survey (NHANES). Fasting plasma glucose is measured only for participants attending the morning fasting examination, whereas demographic and cardiometabolic covariates are available for the full sample. This creates a natural labeled--unlabeled structure and provides a concrete setting in which to examine whether calibration can convert information in the unlabeled covariate distribution into more efficient regression inference.

We use NHANES to estimate the association between fasting plasma glucose and cardiometabolic risk factors. The covariates in the regression model are age, sex, race/ethnicity indicators, body mass index (BMI), systolic blood pressure (SBP), and diastolic blood pressure (DBP). These variables are fully observed for all participants. Fasting glucose is available for $n=2{,}533$ of the $N=6{,}230$ individuals, forming the labeled sample. To mimic a practically relevant semi-supervised scenario with limited labeled data, we randomly retain only 50\% of the originally labeled observations, yielding $n=1{,}266$ labeled and $N-n=3{,}697$ unlabeled units. We compare the proposed ET estimator with the supervised (OLS) estimator, the density-ratio estimator \citep[DRESS;][]{kawakita2013semi}, and the projection-based estimator \citep[PSSE;][]{song2023general}, using $K=3$ folds for cross-fitting.

\noindent\textbf{Results.}
Table~\ref{tab:semi_supervised} shows that generalized entropy calibration yields clear efficiency gains over the supervised estimator across several regression coefficients, often outperforming the competing semi-supervised alternatives as well. The improvement is most visible in the reduction of standard errors and confidence-interval widths, indicating that the unlabeled sample is being used effectively. In several coefficients, ET attains substantial asymptotic relative efficiency gains compared with the supervised estimator, while maintaining stable point estimates. More broadly, the NHANES analysis shows that the calibration framework is not limited to causal weighting problems: the same balance-and-debias logic can be used to strengthen regression inference when responses are available only for a subset of the sample. In this application, the practical value of the framework lies less in reproducing a benchmark than in improving efficiency while preserving a principled correction for selective outcome availability. 

These two analyses highlight complementary aspects of generalized entropy calibration. In the LaLonde benchmark, the method operates as a balancing-weight procedure for causal inference and improves agreement with an experimental reference through closer covariate alignment. In NHANES, it operates as a semi-supervised estimator that converts unlabeled covariate information into more efficient regression inference. The common thread is that both applications are instances of the same partially observed-data template: a full-data estimating equation is reconstructed through calibration constraints that combine outcome-related information with a debiasing correction.

\begin{table*}
\centering
\small
\setlength{\tabcolsep}{6pt}
\caption{Comparison of semi-supervised estimators using a 50\% labeled subsample. Notes: Est = estimates, SE = standard error. CIW = confidence interval width, Proportion=50\%, ARE= estimated asymptotic relative efficiency relative to the supervised estimator, we use K=3 folds for cross-fitting.}

\begin{tabular}{llccccccccccc}
\hline
\textbf{Method} &  & Inter. & age & sexF & rHisp & rWhite & rBlack &rAsian & rOther & BMI & SBP & DBP \\
\hline

Supervised
& Est  & 72.31 & 0.43 & -5.86 & -0.79 & -6.33 & -9.21 & -1.41 & -6.71 & 0.82 & 0.10 & -0.12 \\
& SE   &  7.94 & 0.06 &  1.99 &  4.16 &  3.06 &  3.26 &  3.74 &  4.82 & 0.14 & 0.07 & 0.08 \\
& CIW  & 31.13 & 0.23 &  7.80 & 16.32 & 12.01 & 12.78 & 14.66 & 18.88 & 0.55 & 0.26 & 0.32 \\
\hline

DRESS
& Est  & 71.55 & 0.39 & -4.24 & -2.16 & -6.33 & -7.90 & -2.51 & -6.56 & 0.64 & 0.16 & -0.12 \\
& SE   & 10.86 & 0.07 &  1.98 &  5.28 &  3.61 &  3.64 &  3.83 &  3.77 & 0.17 & 0.11 & 0.06 \\
& CIW  & 42.57 & 0.27 &  7.75 & 20.69 & 14.16 & 14.28 & 15.02 & 14.79 & 0.67 & 0.42 & 0.22 \\
& ARE  &  0.54 & 0.70 &  1.01 &  0.62 &  0.72 &  0.80 &  0.95 &  1.63 & 0.65 & 0.36 & 2.17 \\
\hline

PSSE
& Est  & 71.98 & 0.40 & -4.55 & -2.16 & -6.91 & -8.25 & -2.73 & -7.22 & 0.66 & 0.16 & -0.14 \\
& SE   & 10.80 & 0.07 &  1.97 &  5.25 &  3.60 &  3.63 &  3.82 &  3.76 & 0.17 & 0.11 & 0.06 \\
& CIW  & 42.34 & 0.27 &  7.72 & 20.56 & 14.11 & 14.24 & 14.97 & 14.75 & 0.67 & 0.42 & 0.22 \\
& ARE  &  0.54 & 0.71 &  1.02 &  0.63 &  0.73 &  0.81 &  0.96 &  1.64 & 0.66 & 0.37 & 2.19 \\
\hline

ET
& Est  & 77.08 & 0.40 & -4.97 & -0.03 & -5.21 & -7.55 & -1.70 & -5.21 & 0.67 & 0.12 & -0.15 \\
& SE   &  8.07 & 0.05 &  1.59 &  3.83 &  2.69 &  2.86 &  3.07 &  2.96 & 0.14 & 0.08 & 0.08 \\
& CIW  & 31.63 & 0.20 &  6.24 & 15.02 & 10.55 & 11.22 & 12.05 & 11.59 & 0.53 & 0.31 & 0.30 \\
& ARE  &  0.97 & 1.33 &  1.56 &  1.18 &  1.29 &  1.30 &  1.48 &  2.65 & 1.04 & 0.70 & 1.16 \\
\hline
\end{tabular}
\label{tab:semi_supervised}
\end{table*}

\section{Conclusion}\label{sec:conclusion}

This paper develops a unified calibration-based framework for inference with partially observed data. The central idea is that augmentation can be represented implicitly through calibration: rather than adding an explicit augmentation term to an inverse-probability weighted estimating equation, we construct generalized entropy weights that simultaneously enforce a balancing constraint based on outcome-related information and a debiasing constraint tied to a working propensity-score model. This yields a common inferential template for parameters defined by general estimating equations under missing at random, spanning causal inference, semi-supervised learning, and regression with missing covariates.

The framework has several important consequences. When either the outcome-regression model or the propensity-score model is correctly specified, the resulting estimator is consistent, and when both are correct it attains the semiparametric efficiency bound. More distinctively, Corollary~\ref{cor:var-dom} shows that under correct propensity-score specification the proposed estimator is never less efficient than classical AIPW and is strictly more efficient whenever the debiasing information is not already contained in the augmentation space. In this sense, generalized entropy calibration does more than stabilize weights: it enlarges the effective projection space used for augmentation and thereby improves efficiency under outcome-model misspecification.

The simulations and real-data analyses support this theoretical picture. Across the examples considered in the paper, the proposed estimators recover the efficient benchmark when both working models are correct, retain the expected robustness when only one working model is correct, and show their clearest gains when the propensity-score model is correct but the outcome model is misspecified. The empirical results therefore reinforce the main theoretical message of the paper: calibration can serve not only as a balancing device, but also as a general mechanism for doubly robust and efficient inference with partially observed data.

Several extensions merit further investigation. First, the current framework assumes a missing-at-random mechanism; extending the calibration formulation to nonignorable missingness would substantially broaden its scope. Second, when the calibration function is high-dimensional relative to the sample size, the entropy optimization problem may require regularization or dimension reduction for stable implementation. More broadly, the proposed framework suggests that calibration can provide a common language for weighting, augmentation, and prediction in a wide range of problems involving incomplete or integrated data.

%\section*{Acknowledgement}
%Jae Kwang Kim is the corresponding author of the paper. 

\begin{acks}[Data and Code Availability]
The real data analyses in this study use publicly available data. The National Supported Work (NSW) experimental data \citep{lalonde1986evaluating, dehejia1999causal} are available from the National Bureau of Economic Research (NBER) at \url{http://www.nber.org/~rdehejia/data/} and can be loaded using the \texttt{haven} R package. The PSID and CPS comparison samples are obtained from the \texttt{DAAG} R package. 
The NHANES data are publicly available from the Centers for Disease Control and Prevention (CDC) at \url{https://www.cdc.gov/nchs/nhanes} and can be accessed using the \texttt{nhanesA} R package. 
All simulation and analysis code used to generate the results in this article is available from the authors upon request.
%\textcolor{red}{(We can give the Github link here.) }
\end{acks}

\section*{Disclosure Statement} 

No potential conflict of interest was reported by the author(s).

\begin{supplement}
\stitle{Supplementary Material for ``A Calibration Framework for Inference with Partially Observed Data''.}
\sdescription{The supplementary material contains all regularity conditions (Section~A) and proofs of the theoretical results stated in the main paper (Section~B).}
\end{supplement}

\bibliographystyle{imsart-nameyear}

\bibliography{reference}

\end{document}